\documentclass[12pt,a4paper]{article}
\pdfoutput=1

\usepackage{amsmath,amssymb,bm,multirow,bbold,hyperref,slashed}
\usepackage{graphicx}
\usepackage[american]{babel}
\usepackage{latexsym}
\usepackage{mathrsfs}
\usepackage{amsfonts}
\usepackage{dcolumn} 
\usepackage{float}
\usepackage{cite}
\usepackage{longtable}
\usepackage{bbm}
\usepackage{cite}
\usepackage{subcaption}
\usepackage{standalone}
\usepackage{fixmath}
\usepackage{color}

 \usepackage{feynmf}

\usepackage[utf8]{inputenc}
\usepackage{epsfig}

\newcommand{\lsim}{\stackrel{<}{_\sim}}

\newcommand*{\B}[1]{\ifmmode\bm{#1}\else\textbf{#1}\fi}

\allowdisplaybreaks
\begin{document}

\begin{titlepage}
\begin{flushright}
{
ZU-TH 07/19
}
\end{flushright}
\vskip 1cm

\setcounter{footnote}{0}
\renewcommand{\thefootnote}{\fnsymbol{footnote}}
\vspace*{1cm}
\begin{center}
{\LARGE \bf Scalar-involved three-point Green functions }
\\ [13pt]{\LARGE \bf and their phenomenology}
\vspace{2cm} \\
{\sc  Ling-Yun Dai}$^{1,}$\footnote{Email:~dailingyun@hnu.edu.cn}
{\sc , Javier~Fuentes-Mart\'{\i}n}$^{2,}$\footnote{Email:~fuentes@physik.uzh.ch} and
{\sc  Jorge~Portol\'es}$^{3,}$\footnote{Email:~Jorge.Portoles@ific.uv.es}
\vspace{1.5cm} \\
$^1$School of Physics and Electronics, Hunan University, Changsha 410082, China \\
$^2$Physik-Institut, Universit\"at Z\"urich, CH-8057 Z\"urich, Switzerland \\
$^3$Instituto de F\'{\i}sica Corpuscular, CSIC - Universitat de Val\`encia,
Apt. Correus 22085, E-46071 Val\`encia, Spain \\ [13pt]
\end{center}

\setcounter{footnote}{0}
\renewcommand{\thefootnote}{\arabic{footnote}}
\vspace*{1cm}

\date{\today}

\begin{abstract}
We analyse {within the framework of resonance chiral theory} the $\langle SA_\mu A_\nu \rangle$ and $\langle SV_\mu V_\nu \rangle$ three-point Green functions, where $S$, $A_{\mu} $ and $V_{\mu}$ are short for scalar, axial-vector and vector $SU(3)$ hadronic currents. We construct the necessary Lagrangian such {that} the Green functions fulfill the asymptotic constraints, at large momenta, imposed by QCD at leading order. We study the implications of our results on the spectrum of scalars in the large-$N_C$ limit, and {analyse} their decays.
\end{abstract}
\end{titlepage}

\section{Introduction}\label{s:intro}
Green functions of quantum fields convey all the dynamics of a quantum field theory describing a system of many interacting particles. Their consistent construction in the hadronic low-energy region {(typically $E \ll 1 \, \mbox{GeV}$)}, driven by non-perturbative Quantum Chromodynamics (QCD), can be thoroughly carried out within the model-independent framework of  Chiral Perturbation Theory (ChPT) \cite{Gasser:1983yg,Gasser:1984gg}. {The predictability of this theory is however} spoiled at ${\cal O}(p^4)$ and higher due to our poor knowledge of the chiral low-energy constants. At higher energies, in the hadronic resonances populated domain
($1 \, \mbox{GeV} \lsim E \lsim 2.5 \, \mbox{GeV}$), the construction of the Green functions has been addressed {only} under several specific model-dependent assumptions, {such as} the Extended Nambu-Jona-Lasinio model \cite{Bijnens:1992uz,Bijnens:1993ap,Bijnens:1994ey} and related {ones} \cite{Bijnens:2003rc}. {D}ifferent implementations of large-$N_C$ \cite{tHooft:1973alw,tHooft:1974pnl,Witten:1979kh}: minimal hadronic ansatz \cite{Peris:1998nj,Peris:2000tw,Golterman:2001pj} and
resonance chiral theory (RChT)
\cite{Ecker:1988te,Ecker:1989yg,Moussallam:1997xx,Knecht:2001xc,RuizFemenia:2003hm,Cirigliano:2004ue,Cirigliano:2005xn,Cirigliano:2006hb,Mateu:2007tr,Kampf:2011ty,Roig:2013baa,Husek:2015wta}, have also been explored in the last {decades}. At {even} higher energies ($2.5 \, \mbox{GeV} \lsim E$), {except} where very narrow hadronic resonances arise, perturbative QCD starts to provide a correct description.
\par
It is clear that QCD should rule the dynamics of those Green functions. However, our lack of knowledge of non-perturbative QCD makes that task very difficult and the use of models of QCD becomes necessary. The construction of those models should include chiral symmetry as a feature to be fulfilled in its low-energy domain. {T}he properties of the model at high-energies are more difficult to implement due to hadronization and hence they are not obvious from a Lagrangian point of view. Several works have addressed this problem within RChT \cite{Ecker:1988te}, {which provides a framework for the evaluation of the Green functions in the intermediate energy region.} This is a Lagrangian setting in terms of pseudo-Goldstone {bosons} and resonances (as matter fields) that, by construction, {respect the} chiral symmetry. {As} in ChPT, {this symmetry} provides the structure of the operators but {gives} no information on the coupling constants. However, due to the presence of resonance fields, the Lagrangian has no obvious counting that controls the number of operators and, consequently, some extra features are needed in its application. On one side Green functions are computed using large-$N_C$ premises \cite{Portoles:2010yt}; this translates, essentially, in a loop expansion generated by the Lagrangian. {This is not enough} to limit the number of operators and, in addition, gives no information on the coupling constants. The extra help comes from the assumption that the correlation functions, as given by RChT ($\Pi_{\mbox{\tiny RChT}}$), can be matched, at large momenta, with the known asymptotic behaviour of Green functions and form factors on QCD grounds ($\Pi_{\mbox{\tiny QCD}}$). This sounds feasible as the RChT result (at tree level) and the operator product expansion (OPE), at ${\cal O}(\alpha_S^0)$, generate an expansion in inverse powers of momenta. The method was originally applied to two-point Green functions in Ref.~\cite{Ecker:1989yg} and later to three-point functions \cite{Moussallam:1997xx,Knecht:2001xc} as:
\begin{equation} \label{eq:match}
\lim_{\lambda \rightarrow \infty} \,\Pi^{\alpha_S^0}_{\mbox{\tiny QCD}} (\lambda q)\, = \,
\lim_{\lambda \rightarrow \infty} \,\Pi_{\mbox{\tiny RChT}}^{\mbox{\tiny tree}} (\lambda q) \, .
\end{equation}
Short-distance constraints are also imposed on vertex functions (form factors) by considering their Brodsky-Lepage \cite{Lepage:1980fj} asymptotic behaviour, using parton dynamics \cite{Ecker:1989yg,Pich:2008jm}. These approaches can provide valuable information
on the structure of the operators and their {coupling} constants. Moreover, as the later do not depend on the masses of the pseudo-Goldstone bosons, the procedure can be carried out in the chiral limit. The question of the feasibility of this matching was discussed in Ref.~\cite{Bijnens:2003rc}.
\par
The above-mentioned procedure is particularly transparent for Green functions that are order parameters of the spontaneous breaking of the chiral symmetry, i.e. those that do not receive contributions of perturbative QCD, in the chiral limit, at large momentum transfers and, therefore, show a rather smooth behaviour.  Several works along this line have been produced \cite{Knecht:2001xc,RuizFemenia:2003hm,Cirigliano:2004ue,Cirigliano:2005xn,Cirigliano:2006hb,Mateu:2007tr,Kampf:2011ty,Husek:2015wta} with noticeable results. One of the key
issues in order to carry out the matching procedure in Eq.~(\ref{eq:match}) lies in the construction of the appropriate operators in the RChT Lagrangian that make
the matching possible. The procedure may not always be feasible \cite{Bijnens:2003rc}, but most of the time it is just a matter of looking for the suitable operators.
In Ref.~\cite{Moussallam:1994at} it was pointed out the difficulty involved in the matching for the $\langle SV_{\mu} V_{\nu}  \rangle$ Green function (where $S$ and $V_{\mu}$ are short for scalar and vector QCD currents, respectively) using a Proca representation for the vector resonance fields in RChT. As expected, the authors satisfied the matching by including a higher order (in derivatives) RChT operator that was needed to enforce the QCD short-distance behaviour even though it was non-leading at low energies.
In this article we perform a systematic analysis of the $\langle SV_\mu V_\nu \rangle$ and $\langle S A_{\mu} A_{\nu} \rangle$ Green functions ($A_{\mu}$ is short for axial-vector QCD
current) using an antisymmetric representation for the spin-1 resonances in the RChT framework. We will fulfill the matching indicated by Eq.~(\ref{eq:match}) for both Green functions by constructing a minimal set of RChT operators that provide the correct short-distance behaviour. {We consider} tree-level diagrams only and, accordingly, {work in} the $N_C \rightarrow \infty$ limit. {Moreover we restrict our large-$N_C$ description to} only one multiplet for each hadron type: scalars, vectors and axial-vectors. As a final result we obtain several relations between the relevant coupling constants of the Lagrangian.
\par
The description, classification and dynamics of hadronic scalar meson resonances, with masses $M_S \lsim \, 2 \, \mbox{GeV}$, has a long story of successes and failures (see the corresponding note in Ref.~\cite{Tanabashi:2018oca}). The light-quark spectrum of meson resonances is populated by many scalar states whose identification as $SU(3)$ octets/nonets is far from clear and that are, probably, an admixture of exotic states that involve tetraquarks or even glueballs. The unsolved non-perturbative dynamics does not allow us to identify the nature of the bound states generated by QCD. Experimentally one observes a number of
$J^P=0^+$ states that could fit into two $U(3)$ nonets constituted by quarks. Our present knowledge points out to usual $[\overline{q}q]$ states but also
tetraquark ones $[\overline{q}q][\overline{q}q]$ \cite{Jaffe:2004ph}. The existence of a glueball (with $J^P=0^+$ and of similar properties to the quark resonances) with mass in the upper part of our spectrum ($\sim 2 \, \mbox{GeV}$) was also pointed out some time ago by the lattice \cite{Bali:1993fb,Chen:1994uw}. Hence it is expected that all the scalar resonances in this energy region could be an admixture of all these basic states. 
\par
By construction, the leading multiplets of resonances described by RChT should correspond to those remaining in the $N_C \rightarrow \infty$ limit. However, while this identification does not create discussion for vector, axial-vector and pseudoscalar resonances, the scalar case is much more complex.
In Ref.~\cite{Cirigliano:2003yq} a study within RChT in the large-$N_C$ framework identified the preferred lightest scalar nonet as the one constituted by $S_{\mbox{\tiny $\infty$}} = \{ f_0(980), K^*_0(1430), a_0(1450), f_0(1500) \}$, assuming that the $a_0(980)$ is dynamically generated and making an octet together with $f_0(500)$ and $K^*_0(700)$ as a subleading spectrum.
{In Ref.~\cite{Dai:2017uao}, a new method to study the large-$N_C$ behavior
of the final states interactions (FSI) within the dispersive approach was proposed. The $N_C$ trajectories of the poles suggest that $f_0(980)$ and $f_0(1370)$ should have the $[\overline{q}q]$ component. This is further confirmed in Ref.~\cite{Dai:2018fmx}, by studying the semi-local duality in the large-$N_C$ limit.} Finally, there is also a broad consensus that $S_{\mbox{\tiny $\infty$}}$ corresponds to the $[\overline{q}q]$ structure while the lightest nonet of resonances is constituted by $[\overline{q}q][\overline{q}q]$ \cite{Jaffe:2004ph,Hooft:2008we} (and references therein), with a possible large mixing between them. Even though we basically agree with this description, we will modify it slightly in order to include the $f_0(1370)$ and the $f_0(1710)$, aiming to account for the glueball in our framework.
\par
Although the experimental situation of the scalar decays is rather poor and uncertain \cite{Tanabashi:2018oca}, we intend to analyse the two-pseudoscalar decays of the spectrum of scalars in RChT, i.e. the $S \rightarrow PP$ decays {of} the leading multiplet in the $N_C \rightarrow \infty$ limit. {In doing so, we will use} the minimal set of operators in this framework. We will conclude that meanwhile the short-distance matching procedure of the three-point Green functions requires higher derivative operators in some cases, we do not need {to} introduce subleading operators (in the large-$N_C$ counting) to fulfill the matching. On the contrary, the experimental data on the $S \rightarrow PP$ decays will require to break manifestly that counting by introducing subleading operators. Hence we conclude that the scalar related couplings in the matching of the Green functions are not given by the $N_C \rightarrow \infty$ limit.
\par
In Section 2 we recall the RChT framework within our large-$N_C$ model, leaving for Section 3 the matching procedure for the $\langle SA_\mu
A_\nu \rangle$ and $\langle S V_\mu V_\nu \rangle$ three-point Green functions. Section 4 is devoted to explain the features of our scalar resonance sector and the results of their decays into two pseudoscalar mesons. We establish our conclusions in Section 5. The chiral notation and several analytical expressions on the decays of scalars into other final states are given in the Appendices.

\section{The large-$N_C$ setting: resonance chiral theory} \label{s:rcht}
RChT is a Lagrangian framework that includes the interaction between the chiral pseudoscalar octet of mesons, in ChPT, and the hadron resonances in the energy region up to $\sim 2 \, \mbox{GeV}$. The symmetries driving the operators are both the chiral ($SU(N)_L \otimes SU(N)_R$) and flavour ($SU(N)$) symmetries, for light flavours, $N=2,3$ \cite{Ecker:1988te,Ecker:1989yg,Cirigliano:2006hb}. By construction the RChT method matches the chiral symmetric results at low energies.
Here we only recall the content needed for our present work. We will only consider scalar, vector and axial-vector resonances, and the case with $N=3$ flavours. For a detailed account and notation {we refer the reader to} Refs.~\cite{Cirigliano:2006hb,Portoles:2010yt} and Appendix A.
\par
The RChT framework starts with the leading chiral Lagrangian involving only the octet of pseudoscalar Goldstone bosons (GB) and external currents. It is given by:
\begin{equation} \label{eq:chpt2}
{\cal L}_{(2)}^{\mbox {\tiny GB}} \, =  \, {\cal L}_{(2)}^{\mbox{\tiny ChPT}} \, = \, \frac{F^2}{4} \, \langle \, u_{\mu} \, u^{\mu} \, + \, \chi_{+} \, \rangle \, ,
\end{equation}
where $F$ is the decay constant of the pion in the chiral limit, and the symbol $\langle \cdot \rangle$ stands for the trace in flavour space. This term collects the information on the spontaneous symmetry breaking of the chiral symmetry and coincides with the same order Lagrangian of ChPT.
\par
RChT has no defined parameter (in the Lagrangian) on which to build a qualified counting to establish a classification for the operators. As the integration of the resonances should provide, generically, the ChPT Lagrangian of ${\cal O}(p^n)$, for $n>2$, it has been customary to classify the RChT operators by the order in momenta of the ChPT operators that they were producing upon integration. Therefore the general structure of the operators is ${\cal O} \sim \langle R_1 R_2 ... R_p \chi(p^n) \, \rangle$, with $R_a$ a $U(3)$ nonet of resonance fields, namely $V_{\mu \nu}$ (vector), $A_{\mu \nu}$ (axial-vector) and $S$ (scalar). Notice that we will use the antisymmetric representation for the spin-1 fields
\cite{Kyriakopoulos:1969zm}, given its relevance in the chiral framework \cite{Gasser:1983yg,Ecker:1989yg}. In addition, $\chi(p^n)$ is a tensor (constructed with chiral invariants in terms of the pseudoscalar Goldstone fields and external currents of ChPT)
of $n$ chiral order (see Appendix~\ref{app:rcht}). {The operators} giving the ${\cal O}(p^4)$ {terms in the} chiral Lagrangian are of the type $\langle R_a \, \chi(p^2) \, \rangle$:
\begin{eqnarray} \label{eq:rcht2}
{\cal L}^V_{(2)} \, &=& \, \frac{F_V}{2 \sqrt{2}} \, \langle V^{\mu \nu} \, f_{+ \mu \nu} \, \rangle \, + \, i \frac{G_V}{\sqrt{2}} \, \langle \, V^{\mu \nu} \,
u_{\mu} \, u_{\nu} \, \rangle \, , \nonumber \\
{\cal L}^A_{(2)} \, & = & \frac{F_A}{2 \sqrt{2}} \, \langle \, A^{\mu \nu} \, f_{- \mu \nu} \, \rangle \, , \nonumber \\
{\cal L}^S_{(2)} \, & = & c_d \, \langle \, S \, u^{\mu} \, u_{\mu} \, \rangle \, + \, c_m \, \langle \, S \, \chi_+ \, \rangle \, ,
\end{eqnarray}
where the real couplings: $F_V$, $G_V$, $F_A$, $c_d$ and $c_m$ are, a priori, unknown.
Those generating the ${\cal O}(p^6)$ chiral Lagrangian have been studied in Ref.~\cite{Cirigliano:2006hb} and have the general structures:
$\langle \, R_a \, \chi(p^4) \, \rangle$, $\langle\,  R_a \, R_b \, \chi(p^2) \, \rangle$ and $\langle \, R_a \, R_b \, R_c \, \rangle$. We will collect those of interest for our study in the next section.
\par
It would also be possible to classify the operators {into sets} that provide the correct asymptotic behaviour of definite n-point Green function of QCD currents, that is, the relation in Eq.~(\ref{eq:match}). As has been concluded in previous studies of these Green functions, one starts with the two-point Green function (and related form factors) and determines {the}
appropriate set of operators and relations between couplings. For instance the study of two-point Green functions, with only one multiplet of resonances (single resonance approximation), gives \cite{Weinberg:1967kj,Ecker:1989yg,Golterman:1999au,Jamin:2000wn,Jamin:2001zq,Pich:2002xy}:
\begin{align} \label{eq:cond2point}
F_V \, G_V  & =    F^2 \, ,  &  F_V^2 \, - \, F_A^2  &  =  F^2 \, , &
F_V^2 \, M_V^2  &=   F_A^2 \, M_A^2 \, , \nonumber \\
4 \, c_d \, c_m  &  =  F^2 \, , &  c_d & = c_m \, , 
\end{align}
for the couplings in Eqs.~(\ref{eq:chpt2},\ref{eq:rcht2}). Here $M_V$ and $M_A$ are the masses of the vector and axial-vector nonet, respectively.
When the study is extended to three-point Green functions one may determine an extended set of operators and the initial relations between couplings could be modified \cite{Moussallam:1997xx,Knecht:2001xc,RuizFemenia:2003hm,Cirigliano:2004ue,Cirigliano:2005xn,Cirigliano:2006hb,Mateu:2007tr,Kampf:2011ty,Husek:2015wta}, and so on.
\par
A comment on the nature of the resonances described in the Lagrangian of RChT is needed. This framework is embedded in a large-$N_C$ setting. Accordingly, the spectrum described in the Lagrangian corresponds to states that stay in the $N_C \rightarrow \infty$ limit. Thus our framework cannot contain resonances that are generated by the Lagrangian (for instance on accounts of unitarity) because these are subleading in the $1/N_C$ expansion. A clear case is {the $f_0(500)$}, generated by (or coincident with) a strong $\pi \pi$ wide s-wave.
\par
Together with ${\cal L}_{(2)}^{\mbox{\tiny GB}}$ in Eq.~(\ref{eq:chpt2}) and the Lagrangian involving resonances, RChT {requires the addition of operators with the same structure as the ones in}
the ChPT Lagrangian at {${\cal O}(p^4)$ \cite{Gasser:1984gg}, ${\cal O}(p^6)$ \cite{Bijnens:1999sh},
and so on,} although with different couplings.
 It is well known that the low-energy couplings in ChPT are, at least at
${\cal O}(p^4)$, mostly saturated by the contribution of the lightest multiplets of resonances~\cite{Ecker:1989yg}. At ${\cal O}(p^6)$ the situation is less clear. 
{Since} the couplings are different {from their ChPT counterparts,} we will denote them as $\hat{L_i}$ and $\hat{C_i}$ (for {${\cal L}_{(4)}^{\mbox{\tiny GB}}$} and {${\cal L}_{(6)}^{\mbox{\tiny GB}}$}, respectively):
\begin{equation} \label{eq:newlecs}
 {\cal L}_{(4)}^{\mbox{\tiny GB}} \, = \, \sum_i \, \hat{L}_i \,{\cal O}_{(4)}^i \; , \qquad \qquad \qquad
 {\cal L}_{(6)}^{\mbox{\tiny GB}} \, = \, \sum_i \, \hat{C}_i \,{\cal O}_{(6)}^i \, .
\end{equation}
Notice that {the dimension of the couplings are } $[\hat{L}_i] = E^0$ and $[\hat{C}_i] = E^{-2}$.
\par
In this article we intend to analyse the three-point Green functions $\langle \, S \, V_{\mu} \, V_{\nu} \, \rangle$ and
$\langle \, S \, A_{\mu} \, A_{\nu} \, \rangle$, {imposing the asymptotic behavior in Eq.~(\ref{eq:match}), at leading order in the $1/N_C$ expansion.} In practice {this}
means that we will evaluate the three-point Green functions in RChT with {tree-level diagrams only. For consistency, we should include in our computations an infinite set of resonances.} We do not know how
to do this in a model-independent way. However, {there are good phenomenological reasons that indicate that the lowest mass states (surviving in the $N_C \rightarrow \infty$ limit) contribute dominantly, as has been shown for instance in} the determinations of the ${\cal O}(p^4)$ low-energy couplings \cite{Ecker:1988te}. This is in agreement with the usual decoupling of effective field theories where {the contributions from heavy} mass states to the low-energy theory {is suppressed by powers of} $E/M$, {with} $E$ the energy scale of the effective theory and $M$ the mass of the decoupled state. Accordingly, we {\em model} our $N_C \rightarrow \infty$ setting by including {only} the lightest multiplet of resonances {for each hadron type.}
\par
The identification of the nonets in Eq.~(\ref{eq:rcht2}) is simple for vector states
\cite{Tanabashi:2018oca}: $V_{\mu \nu}(1^-) = \{ \rho(770), K^*(892),\omega(782),\phi(1020)\}$. For axial-vector mesons the situation
is slightly more complicated \cite{Cirigliano:2003yq}:  $A_{\mu \nu}(1^+) = \{ a_0(1260), K_1(1270), f_1(1285), f_1(1420) \}$, since the strange doublet could also be $K_1(1400)$ or an admixture of both. The common feature of these two multiplets is that they correspond to the
lightest states (experimentally identified) with those quantum numbers. For the scalar resonance case (and the glueball) the identification of the lightest nonet, surviving at $N_C \rightarrow \infty$, seems not to concur with the lightest nonet but with one of higher mass. We delay this discussion to Section~4.

\section{Three-point Green functions from RChT} \label{s:three}
{Similarly to the relations in Eq.~(\ref{eq:cond2point}), based on two-point Green functions, one can obtain additional constraints on the RChT couplings by analyzing the three-point Green functions.} A lot of work has already been employed in their study \cite{Moussallam:1997xx,Knecht:2001xc,RuizFemenia:2003hm,Bijnens:2003rc,Cirigliano:2004ue,Cirigliano:2005xn,Cirigliano:2006hb,Mateu:2007tr,Kampf:2011ty,Husek:2015wta}. {Here we focus on the scalar-involved Green functions $\langle \, S \, A_{\mu} \, A_{\nu} \, \rangle$ and $\langle \, S \, V_{\mu}  \, V_{\nu} \, \rangle$.} Both of them are order parameters of the spontaneous chiral symmetry breaking and, consequently, vanish at ${\cal O}(\alpha_S^0)$ in the chiral limit.
\par
\begin{figure}[t]
\begin{center}
\includegraphics[scale=1]{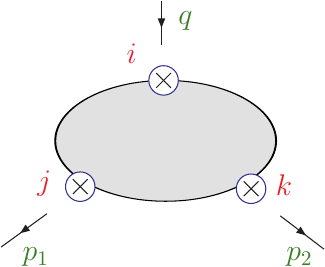}
\caption[]{\label{fig:1} Identification of momenta for the $\Pi^{ijk}_{123}$ Green function. Here $q=p_1+p_2$.}
\end{center}
\end{figure}

The definition of these Green functions is given by
\begin{equation} \label{eq:defGf}
\Pi^{ijk}_{123}(p_1,p_2) \, = \, i^2 \int  d^4x  d^4y \,  e^{i(p_1 \cdot x + p_2 \cdot y)}
\langle 0 | T \left\{ \left( \overline{\psi} \Gamma_1 \frac{\lambda^i}{2} \psi \right) (0)  \left( \overline{\psi} \Gamma_2 \frac{\lambda^j}{2} \psi \right) (x)  \left( \overline{\psi} \Gamma_3 \frac{\lambda^k}{2} \psi \right) (y) \right\} |0 \rangle \, ,
\end{equation}
where $\Gamma_i = 2$ for the scalar current, $\Gamma_i = \gamma_{\mu}$ for the vector current and $\Gamma_i = \gamma_{\mu} \gamma_5$ for the axial-vector current. {Our conventions for the momenta are defined} in Figure~\ref{fig:1}.
We will proceed to determine the general structure of those Green functions as provided by their chiral Ward identities, $SU(3)_V$, parity and time reversal. Then we will obtain their short-distance behaviour at leading order in the momenta expansion. We also calculate their expressions using RChT and including the necessary operators {such that} we have a perfect matching in the momenta expansion, {following the relation in} Eq.~(\ref{eq:match}). A simplifying aspect of the procedure is that, since the couplings do not depend on the masses of the pseudoscalar mesons, we can perform this operation in the chiral limit. {Since our Green functions are order parameters of the chiral symmetry breaking, this implies that there is no perturbative contribution in the parton calculation, at least at ${\cal O}(\alpha_S^0)$.}

\subsection{$\mathbold{\langle \, S \, A_{\mu} \, A_{\nu} \, \rangle}$}
\label{ss:saa}
\begin{figure}[t]
\begin{center}
\includegraphics[scale=0.65]{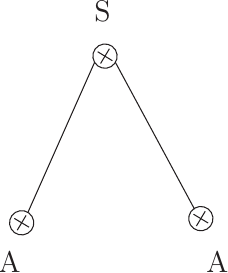}
\caption[]{\label{fig:2} Contribution to $\langle SA_{\mu}A_{\nu} \rangle$ from the chiral Lagrangian ${\cal L}_{(2)}^{\mbox{\tiny ChPT}}$.}
\end{center}
\end{figure}

The $\langle SA_{\mu} A_{\nu} \rangle$ Green function is defined by:
\begin{equation} \label{eq:saagf}
\left( \Pi^{ijk}_{SAA} \right)_{\mu \nu} \, = \, i^2 \, \int d^4x \, d^4y \,
e^{i \left( p_1 \cdot x + p_2 \cdot y \right)} \, \langle 0 | T \left\{ S^i(0) A_{\mu}^j(x) A_{\nu}^k(y) \right\} |
0 \rangle \, ,
\end{equation}
where
\begin{eqnarray} \label{eq:scur}
 S^i(x) \, = \, \left( \bar q \lambda^i q \right)(x) \; \; \;  & , & \; \; \; \,
A^i_{\mu}(x) \, = \, \left( \bar q \gamma_{\mu} \gamma_5 \frac{\lambda^i}{2} q \right)(x) \; ,
\end{eqnarray}
with $q(x) = (u,d,s)^T$ the quark fields.
In $SU(3)$ it satisfies the Ward identities:
\begin{eqnarray} \label{eq:WARD}
 p_1^{\mu} \left( \Pi^{ijk}_{SAA} \right)_{\mu \nu} & = & - 2 \, d^{ijk} B_0 F^2 \, \frac{(p_2)_{\nu}}{p_2^2} \, ,
\nonumber \\
 p_2^{\nu} \left( \Pi^{ijk}_{SAA} \right)_{\mu \nu} & = & - 2 \, d^{ijk} B_0 F^2 \, \frac{(p_1)_{\mu}}{p_1^2} \, .
\end{eqnarray}
Here $B_0$ parameterizes the spontaneous chiral symmetry breaking and it has been defined in Eq.~(\ref{eq:a_chi}).
The general structure of the Green function is given by:
\begin{eqnarray} \label{eq:saastructure}
 \left( \Pi^{ijk}_{SAA} \right)_{\mu \nu} & = &  d^{ijk} B_0 \left[ - 2 \, F^2 \frac{(p_1)_{\mu} (p_2)_{\nu}}{p_1^2 p_2^2} \,
+ \, {\cal F}_A \left(p_1^2,p_2^2,q^2\right) \, P_{\mu \nu} \, +
\, {\cal G}_A \left(p_1^2,p_2^2,q^2\right) \, Q_{\mu \nu} \right] \; , \nonumber  \\
\end{eqnarray}
{with the generic scalar} functions ${\cal F}_A\left(p_1^2,p_2^2,q^2\right)$ and ${\cal G}_A\left(p_1^2,p_2^2,q^2\right)$,
$q^2 = \left( p_1 + p_2 \right)^2$, and where $P_{\mu \nu}$ and $Q_{\mu \nu}$ are the two Lorentz structures
that vanish upon projection with the $(p_{1})_{\mu}$ and $(p_2)_{\nu}$ momenta:
\begin{eqnarray} \label{eq:pqmunu}
P_{\mu \nu} & = & \left(p_2\right)_{\mu} \left(p_1 \right)_{\nu} \, -\, p_1 \cdot p_2 \, g_{\mu \nu} \, , \nonumber
\\
Q_{\mu \nu} & = & p_1^2\left(p_2\right)_{\mu} \left(p_2\right)_{\nu} \, + \,
p_2^2 \left(p_1\right)_{\mu} \left(p_1\right)_{\nu} \, - \,
p_1 \cdot p_2 \left(p_1\right)_{\mu} \left(p_2\right)_{\nu}  \, - \,
p_1^2 \,p_2^2 \,g_{\mu \nu} \, .
\end{eqnarray}
The Ward identities in Eq.~(\ref{eq:WARD}) are also at the origin of the first term of the Green function in Eq.~(\ref{eq:saastructure}). {This term is recovered in RChT}  by the ${\cal O}(p^2)$ ChPT Lagrangian in Eq.~(\ref{eq:chpt2}) through the diagram in Figure~\ref{fig:2}.
\par
The short-distance behaviour of the $\langle SA_{\mu}A_{\nu} \rangle$ function, at leading order in the momenta
expansion, is given by:
\begin{eqnarray} \label{eq:sh1}
 \lim_{\lambda \rightarrow \infty}\left( \Pi^{ijk}_{SAA} \right)_{\mu \nu} \left( \lambda p_1, \lambda p_2 \right)
& = & -2 \, d^{ijk} \, B_0 F^2 \frac{1}{\lambda^2} \, \frac{1}{p_1^2 \, p_2^2 \, q^2} \, \left[ q^2 \left(p_1\right)_{\mu} \left( p_2 \right)_{\nu} \, + \, Q_{\mu \nu} \, - \, p_1 \cdot p_2 \, P_{\mu \nu} \right] \nonumber \\
& & + \, {\cal O}\left( \frac{1}{\lambda^3} \right) \; ,  \\
\label{eq:sh2}
 \lim_{\lambda \rightarrow \infty}\left( \Pi^{ijk}_{SAA} \right)_{\mu \nu} \left( \lambda p_1, p_2 \right) & =&
-2 \, d^{ijk} \, B_0 F^2 \frac{1}{\lambda} \, \frac{\left(p_1\right)_{\mu} \left( p_2 \right)_{\nu}}{p_1^2 p_2^2} \, +
\, {\cal O}\left( \frac{1}{\lambda^2} \right) \; ,\\
\label{eq:sh3}
 \lim_{\lambda \rightarrow \infty}\left( \Pi^{ijk}_{SAA} \right)_{\mu \nu} \left( p_1, \lambda p_2 \right) & =&
-2 \,d^{ijk} \, B_0 F^2 \frac{1}{\lambda} \, \frac{\left(p_1\right)_{\mu} \left( p_2 \right)_{\nu}}{p_1^2 p_2^2} \, +
\, {\cal O}\left( \frac{1}{\lambda^2} \right) \; ,  \\
\label{eq:sh4}
\lim_{\lambda \rightarrow \infty}\left( \Pi^{ijk}_{SAA} \right)_{\mu \nu} \left( \lambda p_1, q-\lambda p_1 \right) & =& {\cal O}\left( \frac{1}{\lambda^2} \right) \, .
\end{eqnarray}
\par
\begin{figure}[h!]
\hspace*{1cm}
   \begin{subfigure}{4cm}
   \centering\includegraphics[scale=0.65]{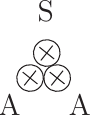}
   \end{subfigure}
   \hspace*{1cm}
   \begin{subfigure}{4cm}
   \centering\includegraphics[scale=0.65]{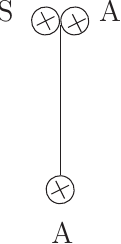}
   \end{subfigure}
   \hspace*{1cm}
   \begin{subfigure}{4cm}
   \centering\includegraphics[scale=0.65]{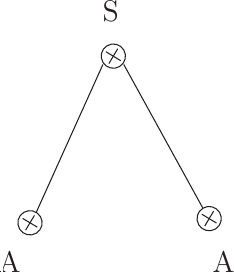}
   \end{subfigure}
   \caption{\label{fig:3} Goldstone boson contributions to the $\langle S A_{\mu} A_{\nu} \rangle$ Green function from the higher-order $\mbox{GB}$ chiral Lagrangian at ${\cal O}(p^4)$ and ${\cal O}(p^6)$.}
\end{figure}

Let us {now} compute ${\cal F}_A$ and ${\cal G}_A$ in RChT at tree level. The content of the Lagrangian, as explained in
Section~\ref{s:rcht} {presents} two main {parts}: the operators with Goldstone boson fields only (and external currents) and those with interactions among them and resonance fields. We have:
\begin{equation} \label{eq:lagsaa}
{\cal L}_{\mbox{\tiny SAA}} \, =  {\cal L}_{(2)}^{\mbox{\tiny GB}} \, + \, {\cal L}_{(4)}^{\mbox{\tiny GB}} \, + \, {\cal L}_{(6)}^{\mbox{\tiny GB}} \, + \, {\cal L}_{(2)}^A \, +
\, {\cal L}_{(2)}^S \, + \, {\cal L}_{A} \, ,
\end{equation}
where the $\mbox{GB}$ Lagrangians have been defined in Eqs.~(\ref{eq:chpt2},\ref{eq:newlecs}). {For the reader's convenience we list the relevant operators in Table~\ref{tab:1}.}
Their contribution to the Green functions are given by the diagrams in Figure~\ref{fig:3}.
\par
\begin{figure}[!h]
   \begin{subfigure}{4cm}
   \centering\includegraphics[scale=0.65]{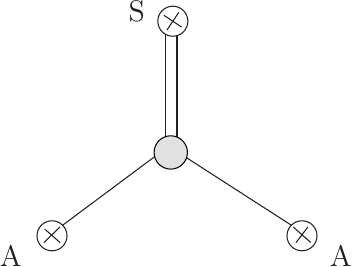}
   \end{subfigure}
   \begin{subfigure}{4cm}
   \centering\includegraphics[scale=0.65]{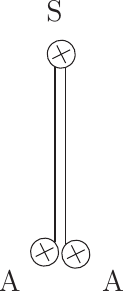}
   \end{subfigure}
   \begin{subfigure}{4cm}
   \centering\includegraphics[scale=0.65]{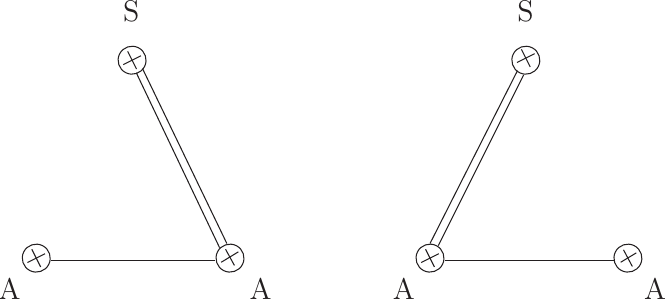}
   \end{subfigure}

   \vspace*{0.6cm}
   \begin{subfigure}{8.5cm}
   \centering\includegraphics[scale=0.65]{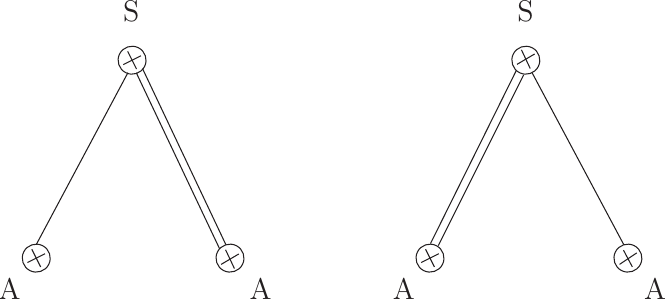}
   \end{subfigure}
   \begin{subfigure}{6cm}
   \centering\includegraphics[scale=0.65]{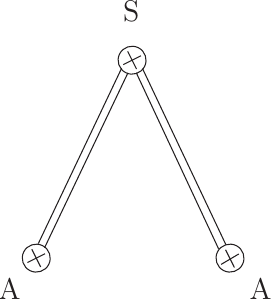}
   \end{subfigure}
   \begin{subfigure}{2cm}
   \centering\includegraphics[scale=0.65]{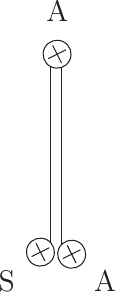}
   \end{subfigure}

   \vspace*{0.6cm}
   \hspace*{1cm}
   \begin{subfigure}{8.5cm}
   \centering\includegraphics[scale=0.65]{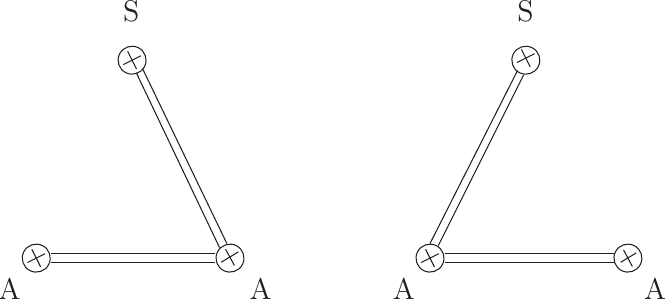}
   \end{subfigure}
   \hspace*{0.9cm}
   \begin{subfigure}{6cm}
   \centering\includegraphics[scale=0.65]{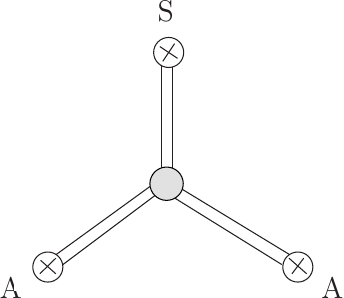}
   \end{subfigure}

   \vspace*{0.6cm}
   \hspace*{3.5cm}
   \begin{subfigure}{8cm}
   \centering\includegraphics[scale=0.65]{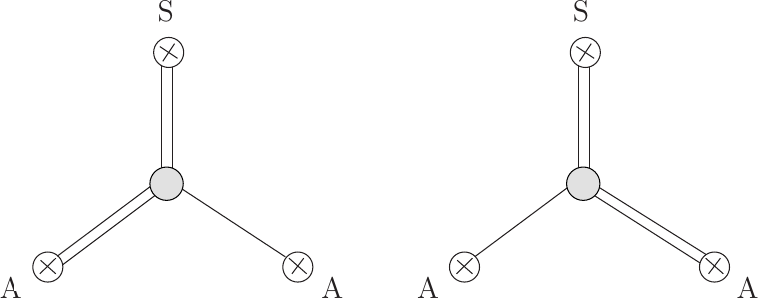}
   \end{subfigure}
   \caption{\label{fig:4} Diagrams contributing to the $\langle S A_{\mu} A_{\nu} \rangle$ Green function in RChT. Goldstone bosons and resonance states are represented by single and double lines, respectively. }
\end{figure}

Next we consider the resonance contributions. {The Lagrangians} ${\cal L}_{(2)}^A$ and ${\cal L}_{(2)}^S$ are given in Eq.~(\ref{eq:rcht2}) while in ${\cal L}_A$ we include those
operators with resonances, Goldstone fields and external currents that, upon integration of the resonances, originate the ${\cal O}(p^6)$ ChPT Lagrangian. They have been constructed in Ref.~\cite{Cirigliano:2006hb}. Those contributing to our Green function are also collected in Table~\ref{tab:1}. They {contribute through} the diagrams in Figure~\ref{fig:4}. Previous short-distance constraints already concluded that $\lambda_{17}^S = \lambda_{18}^S = \lambda_{17}^A=0$~\cite{Cirigliano:2006hb}. We {include these couplings in} our analysis and {we set them to zero at the very end}.
\begin{table}[!t]
\begin{center}
\renewcommand{\arraystretch}{1.5}
\begin{tabular}{|c|c||c|c||c|c|}
\hline
\multicolumn{1}{|c}{Coupling} &
\multicolumn{1}{|c||}{Operator}  &
\multicolumn{1}{c}{Coupling} &
\multicolumn{1}{|c||}{Operator}  &
\multicolumn{1}{c}{Coupling} &
\multicolumn{1}{|c|}{Operator}   \\
\hline
\hline
 $F^2/4$ & $\langle u_{\mu}\,u^{\mu}\,+\,\chi_+  \, \rangle$ & $\lambda_{12}^S$ & $\langle \, S \, \{ \, \nabla_{\alpha} \, f_-^{\mu \alpha}, \, u_{\mu} \, \} \, \rangle$ & $\lambda_1^{SA}$  &
$\langle \, \{ \, \nabla_{\mu} \, S, \, A^{\mu \nu} \, \} \, u_{\nu} \, \rangle$ \\
$\tilde{L}_{5}$ &  $\langle\, u_{\mu}\, u^{\mu}\, \chi_+\, \rangle$ & $\lambda_{16}^S$ & $\langle \, S \, f_{- \, \mu \nu} \, f_{-}^{\mu \nu} \, \rangle$ & $ \lambda_2^{SA}$ & $ \langle \, \{ \, S , \, A_{\mu \nu} \, \} f_{-}^{\mu \nu} \, \rangle$ \\
$\tilde{C}_{12}$ & $\langle\, h_{\mu\nu}\,h^{\mu\nu}\,\chi_+\, \rangle$  & $\lambda_{17}^S$ & $\langle \, S \, \nabla_{\alpha} \, \nabla^{\alpha} ( u_{\mu} \, u^{\mu} ) \, \rangle$ & $\lambda_6^{AA}$ & $\langle \, A_{\mu \nu} \, A^{\mu \nu} \, \chi_+ \, \rangle $ \\
$\tilde{C}_{80}$  & $\langle\, f_{-\mu\nu}\, f_-^{\mu\nu}\, \chi_+ \,\rangle$ & $\lambda_{18}^S$ & $\langle \, S \, \nabla_{\mu} \, \nabla^{\mu} \, \chi_{+} \, \rangle$ & $\lambda^{SAA}$ & $ \langle \, S \, A_{\mu \nu} \, A^{\mu \nu} \, \rangle$ \\
$\tilde{C}_{85}$ & $\langle\, f_{-\mu\nu}\,\{ \chi_+^\mu,\,u^\nu\}\, \rangle$  & $\lambda_6^A$ & $ \langle \, A_{\mu \nu} \, [ \, u^{\mu} \, , \nabla^{\nu} \, \chi_+ ] \, \rangle$ & & \\
 &  & $\lambda_{16}^A$ & $ \langle \, A_{\mu \nu} \, \{ \, f_{-}^{\mu \nu}, \, \chi_{+}\} \, \rangle $ & & \\
 &  & $\lambda_{17}^A$ & $\langle \, A_{\mu \nu} \, \nabla_{\alpha} \nabla^{\alpha} f_{-}^{\mu \nu} \, \rangle $ & & \\
\hline
\multicolumn{4}{c}{}
\end{tabular}
\end{center}
\vspace*{-1cm}
\caption{\label{tab:1}
Couplings and operators in ${\cal L}_{\mbox{\tiny{SAA}}}$ contributing to the $\langle S A_{\mu} A_{\nu} \rangle$ Green function. Those with resonances are collected from Ref.~\cite{Cirigliano:2006hb}. On the left two columns we collect the operators with only Goldstone bosons given by ChPT. On the middle two columns we collect the operators with one resonance $\langle R \chi(p^4) \rangle$. On the right two columns we list the operators with more than one resonance: $\langle RR \chi(p^2)\rangle$ and $\langle RRR \rangle$. Note that the dimensions of these couplings are $[\lambda_i^R] = E^{-1}$, $[\lambda_i^{RR}] = E^0$ and $[\lambda^{SAA}] = E$.}
\end{table}
\par
The final result for the ${\cal F}_A$ and ${\cal G}_A$ functions defined in Eq.~(\ref{eq:saastructure}) is:
\begin{eqnarray} \label{eq:fsaa}
{\cal F}_A(p_1^2,p_2^2,q^2) & = & 32 \, \left(\hat{C}_{12} \, - \, \hat{C}_{80} \, - \, \hat{C}_{85} \right) \,  - \, 32 \,  \lambda_{16}^S \, P_S \, - \, 16 \, \lambda_6^{AA} \, P_A(p_1^2) \, P_A(p_2^2) \, \,  \nonumber \\
& & + \, 8 \, \sqrt{2} \, \left( 2 \,  \lambda_{16}^A \, - \, \lambda_6^A \, + \, ( \, \lambda_1^{SA} \, + \,  2 \, \lambda_2^{SA} \, ) \,P_S  \, \right)\, \left( P_A(p_1^2) + P_A(p_2^2) \right) \nonumber \\
& &  - \, 16 \, \lambda^{SAA} \, P_S \, P_A(p_1^2) \, P_A(p_2^2) \,,
\end{eqnarray}
and
\begin{eqnarray} \label{eq:gsaa}
{\cal G}_A(p_1^2,p_2^2,q^2) & = & \frac{8}{p_1^2 \,p_2^2} \bigg( \, 2 \, \hat{L}_5 \, + \, 4 \, \hat{C}_{12} (\, p_1^2 \, + \, p_2^2 \, - \, q^2 \, ) \, - \, 2 \, \hat{C}_{85} \,  (\, p_1^2 \, + \, p_2^2 \, ) \, + \, 2 \, c_d \, P_S \,  \nonumber \\
& & \qquad \; \; \;  -  \, 2 \,  \lambda_{12}^S \, (p_1^2 \, + \, p_2^2 \, ) P_S \, -  \, 2 \,  \lambda_{17}^S \, q^2 \, P_S   \, \nonumber \\
& &  \qquad \; \; \; - \,  \sqrt{2} \, \left( \lambda_{6}^A \, - \, \lambda_1^{SA} \, P_S \right) \left( p_1^2 \,   P_A(p_1^2) \, + \, p_2^2 \,  P_A(p_2^2) \right) \bigg) \, ,
\end{eqnarray}
where
\begin{eqnarray}  \label{eq:polos}
P_S \,  =   \, \frac{c_m \, - \, \lambda_{18}^S \, q^2}{M_S^2 \, -  \, q^2} \, , \qquad && \qquad \qquad
P_A(p^2) \,  =  \, \frac{F_A \, - \, 2 \, \sqrt{2} \, \lambda_{17}^A \, p^2}{M_A^2 \, - \, p^2} \,  ,
\end{eqnarray}
and $M_S$ and $M_A$ are the mases of the nonet of scalars and axial-vector mesons in the $U(3)$ and chiral limits.
\par
We can now expand our RChT results for the ${\cal F}_A$ and ${\cal G}_A$ functions and impose the constraints by Eqs.~(\ref{eq:sh1},\ref{eq:sh2},\ref{eq:sh3}) and  (\ref{eq:sh4}). We get:
\begin{eqnarray} \label{eq:saares}
\hat{L}_{5} \! \! &=& \! \! \hat{C}_{12} \, = \, \hat{C}_{80} \, = \, \hat{C}_{85} \, = \, 0 \, , \nonumber \\
\lambda_6^A  \! \! &=&  \! \! \lambda_{16}^A \, = \, \lambda_{12}^S \, = \, \lambda_{16}^S \, =  \, 0 \, , \nonumber \\
\lambda_6^{AA} \! \! &=&  \! \! - \, \frac{F^2}{16 \, F_A^2} \, , \nonumber \\
\lambda_1^{SA} \! \! &=&  \! \! \frac{1}{\sqrt{2} \, F_A} \, \left( c_d \, - \, \frac{F^2}{8 \, c_m} \right) \, , \nonumber \\
\lambda_2^{SA} \! \! &=&  \! \! - \, \frac{c_d}{2 \, \sqrt{2} \, F_A} \, .
\end{eqnarray}
It is interesting to observe that the low-energy couplings of the $\mbox{GB}$ Lagrangians vanish. This strengthens the notion of resonance dominance of the chiral couplings.

\subsection{$\mathbold{\langle \, S \, V_{\mu} \, V_{\nu} \, \rangle}$}
\label{ss:svv}
We proceed analogously with the $\langle S V_{\mu}  V_{\nu} \rangle$ Green function defined by:
\begin{equation} \label{eq:svvgf}
\left( \Pi^{ijk}_{SVV} \right)_{\mu \nu} \, = \, i^2 \, \int d^4x \, d^4y \,
e^{i \left( p_1 \cdot x + p_2 \cdot y \right)} \, \langle 0 | T \left\{ S^i(0) V_{\mu}^j(x) V_{\nu}^k(y) \right\} |
0 \rangle \, ,
\end{equation}
where
\begin{eqnarray} \label{eq:vectorc}
V^i_{\mu}(x) \, = \, \left( \bar q \gamma_{\mu} \frac{\lambda^i}{2} q \right)(x) \;,
\end{eqnarray}
and the scalar current as defined in Eq.~(\ref{eq:scur}).
In the $SU(3)$ limit it satisfies the Ward identities:
\begin{eqnarray} \label{eq:wardsvv}
 p_1^{\mu} \left( \Pi^{ijk}_{SVV} \right)_{\mu \nu} & = & 0 \, ,
\nonumber \\
 p_2^{\nu} \left( \Pi^{ijk}_{SVV} \right)_{\mu \nu} & = & 0\, .
\end{eqnarray}
Its general structure is given by:
\begin{eqnarray} \label{eq:svvstructure}
 \left( \Pi^{ijk}_{SVV} \right)_{\mu \nu} & = &  d^{ijk} B_0 \left[
 \, {\cal F}_V\left(p_1^2,p_2^2,q^2\right) \, P_{\mu \nu} \, +
\, {\cal G}_V\left(p_1^2,p_2^2,q^2\right) \, Q_{\mu \nu} \right] \; ,
\end{eqnarray}
where $P_{\mu \nu}$ and $Q_{\mu \nu}$ have been defined in Eq.~(\ref{eq:pqmunu}).
\par
The short-distance behaviour of the $\langle SVV \rangle$ function, at leading order in the momenta
expansion, reads \footnote{It is possible to vary the high energy behavior of the Green function as
\begin{eqnarray}
 \lim_{\lambda_1, \lambda_2 \rightarrow \infty}\left( \Pi^{ijk}_{SVV} \right)_{\mu \nu} \left( \lambda_1 p_1, \lambda_2 p_2 \right)
\;. \nonumber
\end{eqnarray}
Since $\lambda_1,\lambda_2$ arbitrarily go to infinity, the matching in the short distance region should be fulfilled for each momentum independently.}:
\begin{eqnarray} \label{eq:sh1v}
 \lim_{\lambda \rightarrow \infty}\left( \Pi^{ijk}_{SVV} \right)_{\mu \nu} \left( \lambda p_1, \lambda p_2 \right)
& = & -  \, d^{ijk} \, B_0 F^2 \frac{1}{\lambda^2} \, \frac{1}{p_1^2 \, p_2^2 \, q^2} \, \left[ 2 \, Q_{\mu \nu} \, + \, \left( p_1^2 + p_2^2 + q^2 \right) P_{\mu \nu} \right] \, \nonumber \\
& & + \,  {\cal O}\left( \frac{1}{\lambda^3} \right) \; ,  \\
\label{eq:sh2v}
 \lim_{\lambda \rightarrow \infty}\left( \Pi^{ijk}_{SVV} \right)_{\mu \nu} \left( \lambda p_1, p_2 \right) & =&
-2 \, d^{ijk} \,  \frac{1}{\lambda} \, \frac{\Pi_{VT}(p_2^2)}{p_1^2} \, P_{\mu \nu} \, +
\, {\cal O}\left( \frac{1}{\lambda^2} \right) \; ,\\
\label{eq:sh3v}
 \lim_{\lambda \rightarrow \infty}\left( \Pi^{ijk}_{SVV} \right)_{\mu \nu} \left( p_1, \lambda p_2 \right) & =&
-2 \,d^{ijk} \,\frac{1}{\lambda} \, \frac{\Pi_{VT}(p_1^2)}{p_2^2} \, P_{\mu \nu}  \, +
\, {\cal O}\left( \frac{1}{\lambda^2} \right) \; ,  \\
\label{eq:sh4v}
\lim_{\lambda \rightarrow \infty}\left( \Pi^{ijk}_{SVV} \right)_{\mu \nu} \left( \lambda p_1, q-\lambda p_1 \right) & =& {\cal O}\left( \frac{1}{\lambda^2} \right) \, ,
\end{eqnarray}
and $\Pi_{VT}(p^2)$ is defined by:
\begin{equation} \label{eq:pivvdef}
 \left( p_{\rho} g_{\mu \sigma} \, - \, p_{\sigma} g_{\mu \rho} \right) \, \delta^{ij} \, \Pi_{VT}(p^2) \, = \,
 \int d^4x \, e^{ip\cdot x} \langle 0 | \, T \left\{ V_{\mu}^i (x) \left( \bar q \, \sigma_{\rho \sigma} \,
\frac{\lambda^j}{2} \, q \right) (0) \right\} \, | 0 \rangle \, .
\end{equation}
Let us compute now the ${\cal F}_V$ and ${\cal G}_V$ functions (\ref{eq:svvstructure}) in the RChT formalism. Analogously to the
previous Green function we denote our Lagrangian as:
\begin{equation} \label{eq:lagsvv}
{\cal L}_{\mbox{\tiny SVV}} \, =  \,  {\cal L}_{(2)}^{\mbox{\tiny GB}} \, + \, {\cal L}_{(4)}^{\mbox{\tiny GB}} \,+  \, {\cal L}_{(6)}^{\mbox{\tiny GB}} \, + \, {\cal L}_{(2)}^V \, +
\, {\cal L}_{(2)}^S \, + \, {\cal L}_{V} \, ,
\end{equation}
where ${\cal L}_{(2)}^{\mbox{\tiny GB}}$ is defined in Eq.~(\ref{eq:chpt2}), ${\cal L}_{(4)}^{\mbox{\tiny GB}}$ and ${\cal L}_{(6)}^{\mbox{\tiny GB}}$ are defined in Eq.~(\ref{eq:newlecs}), ${\cal L}_{(2)}^V$ and ${\cal L}_{(2)}^S$ are specified in Eq.~(\ref{eq:rcht2}) and ${\cal L}_V$ includes interaction terms between scalar, vector resonances, and external currents. There is a key difference between the operators needed to match the Green function in the $\langle S A_{\mu} A_{\nu} \rangle$ case and the present ones. The Lagrangian ${\cal L}_A$ only includes  those operators {that}, upon integration of the resonance, contributes to the ChPT ${\cal O}(p^6)$ Lagrangian. Contrary to the $\langle S A_{\mu} A_{\nu} \rangle$ case, these operators are not enough to achieve the matching in the $\langle S V_{\mu} V_{\nu} \rangle$ case. More precisely, if we only include the operators in Table~\ref{tab:2} we would get ${\cal G}_V (p_1^2,p_2^2,q^2) = 0$ and, therefore, we would not be able to fulfill the matching. We thus need {to include additional operators that are listed in} Table~{\ref{tab:3}. They have the chiral structure: $\langle R \chi(p^6) \rangle$, $\langle RR \chi(p^4) \rangle$ and $\langle RRR \chi(p^2) \rangle$ and {yield contributions to both} ${\cal F}_V$ and ${\cal G}_V$.
\begin{table}[!t]
\begin{center}
\renewcommand{\arraystretch}{1.5}
\begin{tabular}{|c|c||c|c|}
\hline
\multicolumn{1}{|c|}{Coupling} &
\multicolumn{1}{c||}{Operator} &
\multicolumn{1}{c|}{Coupling} &
\multicolumn{1}{c|}{Operator} \\
\hline
\hline
$\tilde{C}_{61}$ & $\langle  \, f_{+ \, \mu \nu} \, f_{+}^{\mu \nu} \, \chi_+ \, \rangle$ &  & \\
\hline
$\lambda_{15}^S$ & $\langle \, S \, f_{+ \, \mu \nu} \, f_{+}^{\mu \nu} \,  \rangle$ & $\lambda_6^{VV}$ &
$\langle \,  V_{\mu \nu} \, V^{\mu \nu} \, \chi_{+} \,  \rangle$  \\
$\lambda_6^V$ & $\langle \, V_{\mu \nu} \, \{ f_{+}^{\mu \nu}, \, \chi_{+} \, \} \,  \rangle$ & $\lambda_3^{SV}$ &
$\langle \, \{ \, S, \, V_{\mu \nu} \, \} f_{+}^{\mu \nu} \, \rangle$ \\
$\lambda_{22}^{V}$ & $\langle \, V_{\mu \nu} \, \nabla_{\alpha} \, \nabla^{\alpha} \, f_{+}^{\mu \nu} \, \rangle$ & $\lambda^{SVV}$ &
$\langle  \, S \, V_{\mu \nu} \, V^{\mu \nu} \, \rangle$ \\
\hline
\multicolumn{4}{c}{}
\end{tabular}
\end{center}
\vspace*{-1cm}
\caption{\label{tab:2}
{Operators of ${\cal O}(p^6)$ in ChPT and operators} in ${\cal L}_V$ that, upon integration of the resonances, give chiral operators of ${\cal O}(p^6)$. Short-distance constraints \cite{Cirigliano:2006hb} require that $\lambda_{22}^V = 0$. Note that the dimensions of these couplings is $[\tilde{C}_i] = E^{-2}$, $[\lambda_i^R] = E^{-1}$,
$[\lambda_i^{RR}] = E^0$ and $[\lambda^{SVV}] = E$.}
\end{table}
\begin{table}[!t]
{\footnotesize
\begin{center}
\renewcommand{\arraystretch}{1.5}
\begin{tabular}{|c|c||c|c||c|c|}
\hline
Coupling & Operator & Coupling & Operator& Coupling & Operator  \\
\hline
\hline
$\kappa_1^{SVV}$ & $\langle \nabla^{\mu} V_{\mu \nu} \nabla_{\alpha} V^{\alpha \nu} S \rangle$ &
  $\kappa_1^{SV}$ & $\langle \left\{  V_{\alpha \nu}, \nabla_{\mu} f_{+}^{\mu \nu} \right\} \nabla^{\alpha} S \rangle$ &
     $\kappa_1^{V}$ & $\langle \left\{ \nabla^{\alpha} V_{\alpha \nu},  f_{+}^{\mu \nu} \right\} \nabla_{\mu} \chi_+ \rangle$  \\
$\kappa_2^{SVV}$ & $\langle \left\{ \nabla^{\mu} V_{\mu \nu}, V^{\alpha \nu} \right\} \nabla_{\alpha} S \rangle$ &
  $\kappa_2^{SV}$ & $\langle \left\{ \nabla^{\alpha} V_{\alpha \nu}, \nabla_{\mu} f_{+}^{\mu \nu} \right\} S \rangle$  &
     $\kappa_2^{V}$ & $\langle \left\{ \nabla^{\alpha} V_{\alpha \nu}, \nabla_{\mu} f_{+}^{\mu \nu} \right\} \chi_+ \rangle$ \\
$\kappa_3^{SVV}$ & $\langle \nabla_{\alpha}V_{\mu \nu} \nabla^{\alpha} V^{\mu \nu} S \rangle$ &
  $\kappa_{3}^{SV}$ & $\langle  \left\{\nabla^{\alpha} V_{\mu \nu},  f_{+}^{\mu \nu}\right\}  \nabla_{\alpha} S \rangle$  &
     $\kappa_3^{V}$ & $\langle  \left\{\nabla^{\alpha} V_{\mu \nu},  f_{+}^{\mu \nu}\right\}  \nabla_{\alpha} \chi_+  \rangle$  \\
$\kappa_4^{SVV}$ & $\langle \left\{ \nabla^{\alpha} V_{\mu \nu},  V^{\mu \nu} \right\} \nabla_{\alpha} S \rangle$ &
  $\kappa_{4}^{SV}$ &  $\langle  \left\{\nabla_{\mu} V_{\alpha \nu},  \nabla^{\alpha}f_{+}^{\mu \nu}\right\}   S \rangle$  &
     $\kappa_4^{V}$ &  $\langle  \left\{\nabla_{\mu} V_{\alpha \nu},  \nabla^{\alpha}f_{+}^{\mu \nu}\right\}   \chi_+  \rangle$  \\
$\kappa_5^{SVV}$ & $\langle \left\{ \nabla_{\alpha} V_{\mu \nu}, V^{\alpha \mu} \right\} \nabla^{\nu} S \rangle$ &
  $\kappa_{5}^{SV}$  &  $\langle  \left\{ V^{\alpha \nu},  \nabla_{\alpha}f_{+ \, \mu \nu}\right\} \nabla^{\mu}S \rangle$  &
     $\kappa_5^{V}$  &  $\langle  \left\{\nabla^{\mu} V^{\alpha \nu}, f_{+ \, \mu \nu}\right\} \nabla_{\alpha} \chi_+  \rangle$  \\
$\kappa_6^{SVV}$ & $\langle  \nabla_{\alpha} V_{\mu \nu} \nabla^{\mu} V^{\alpha \nu}  S \rangle$ &
  $\kappa_{1}^{S}$ &  $\langle  \nabla^{\alpha} f_{+}^{\mu\nu} \nabla_{\mu} f_{+ \, \alpha \nu}  S \rangle$  &
     $\kappa_{1}^{VV}$ &  $\langle  \nabla^{\alpha} V^{\mu\nu} \nabla_{\mu} V_{\alpha \nu}  \chi_+ \rangle$  \\
&&$\kappa_{2}^{S}$ & $\langle  \left\{ f_{+}^{\mu\nu}, \nabla^{\alpha} f_{+ \, \alpha \nu}\right\} \nabla_{\mu} S \rangle$  &
     $\kappa_{2}^{VV}$ & $\langle  \left\{ V^{\mu\nu}, \nabla^{\alpha} V_{\alpha \nu}\right\} \nabla_{\mu} \chi_+ \rangle$  \\
&&$\kappa_{3}^{S}$ & $\langle  \nabla^{\alpha} f_{+}^{\mu\nu} \nabla_{\alpha} f_{+ \, \mu\nu}  S \rangle$   &
     $\kappa_{3}^{VV}$ & $\langle  \nabla^{\alpha} V^{\mu\nu} \nabla_{\alpha} V_{\mu\nu}  \chi_+ \rangle$    \\
\hline
\multicolumn{6}{c}{}
\end{tabular}
\end{center}
}
\vspace*{-1cm}
\caption{\label{tab:3}
Operators in ${\cal L}_V$ that, upon integration of the resonances, give chiral operators of ${\cal O}(p^n)$ with $n>6$. The dimensions of the couplings are: $[\kappa_i^{SVV}] = E^{-1}$, $[\kappa_i^{SV,VV}] = E^{-2}$ and $[\kappa_i^{S,V}] = E^{-3}$.}
\end{table}
\par
The complete set of diagrams contributing to $\langle S V_{\mu} V_{\nu} \rangle$ {is} given in Figure~\ref{fig:5}.
\begin{figure}[!h]
   \hspace*{1cm}
   \begin{subfigure}{3.5cm}
   \centering\includegraphics[scale=0.65]{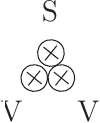}
   \end{subfigure}
   \begin{subfigure}{3.5cm}
   \centering\includegraphics[scale=0.65]{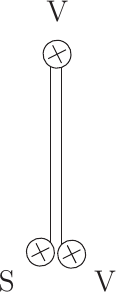}
   \end{subfigure}
   \begin{subfigure}{3.5cm}
   \centering\includegraphics[scale=0.65]{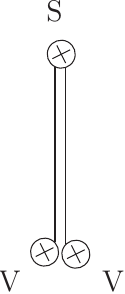}
   \end{subfigure}
   \begin{subfigure}{3.5cm}
   \centering\includegraphics[scale=0.65]{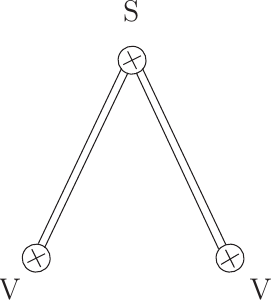}
   \end{subfigure}

   \vspace*{0.6cm}
   \hspace*{2cm}
   \begin{subfigure}{4cm}
   \centering\includegraphics[scale=0.65]{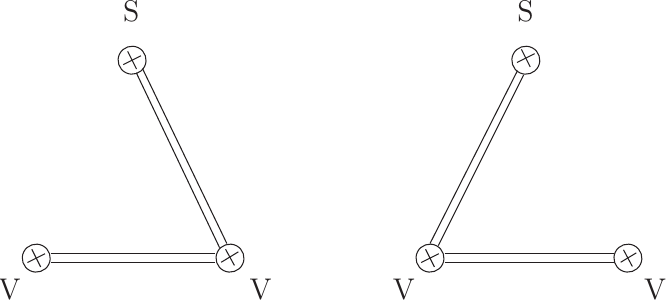}
   \end{subfigure}
   \hspace*{5cm}
   \begin{subfigure}{4cm}
   \centering\includegraphics[scale=0.65]{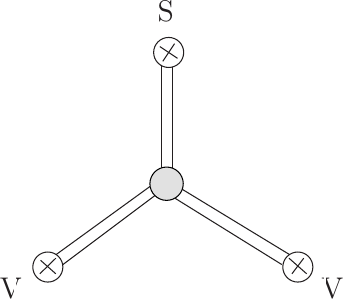}
   \end{subfigure}
   \caption{\label{fig:5} Diagrams contributing to the $\langle S V_{\mu} V_{\nu} \rangle$ Green function in RChT. Goldstone bosons and resonance states are represented by single and double lines, respectively. }
\end{figure}
The resulting expressions for the ${\cal F}_V$ and ${\cal G}_V$ functions are:
\begin{eqnarray} \label{eq:fsvv}
{\cal F}_V(p_1^2,p_2^2,q^2) & = & - \, 32 \, \hat{C}_{61} \, - \, 32 \, \lambda_{15}^S \, P_S \, + \, 16 \,  \sqrt{2} \, \left( \lambda_6^V \, + \lambda_3^{SV} \, P_S \, \right) ( P_V(p_1^2) \, + \, P_V(p_2^2) \, ) \, \nonumber \\
& & -  \, 16 \, \left( \lambda_6^{VV} \,+ \,  \lambda^{SVV} \, P_S \, \right) \,  P_V(p_1^2) \, P_V(p_2^2) \, \nonumber \\
& & -  \, 4 \, \bigg( \left( 2 \, \kappa_2^{SVV} \, + \, 2 \, \kappa_3^{SVV} \, + \kappa_6^{SVV} \right) (\, p_1^2 \, + \, p_2^2 \,) \, \nonumber \\
& &  \qquad \; -  \, \left( 2 \, \kappa_3^{SVV} \, - \, 4 \, \kappa_4^{SVV}  \, + \, 2 \, \kappa_5^{SVV} \, + \, \kappa_6^{SVV} \, \right) \, q^2
\,   \bigg) \, P_S \, P_V(p_1^2) \, P_V(p_2^2) \, \nonumber \\
& & + \, 4 \sqrt{2} \, \left( 2\kappa_1^{SV}-2\kappa_3^{SV}+\kappa_4^{SV}+{\kappa_5^{SV}}\right) \, P_S \, \left( p_1^2 \, P_V(p_2^2) \, + \, p_2^2 \, P_V(p_1^2) \, \right) \,  \nonumber \\
& & + \, 4 \sqrt{2} \, \left( 2\kappa_3^{SV}-\kappa_4^{SV}+{\kappa_5^{SV}}\right)q^2 \, P_S \, \left(P_V(p_1^2) \, + \, P_V(p_2^2) \, \right) \,  \nonumber \\
& & + \, 4 \sqrt{2} \, \left( 2\kappa_3^{SV}+\kappa_4^{SV}-{\kappa_5^{SV}}\right)  P_S  \left(p_1^2\,P_V(p_1^2) \, + \,p_2^2\, P_V(p_2^2) \, \right) \,  \nonumber \\
& & + \, 8  \,
\left(\kappa_1^S+2\kappa_3^S \right) q^2 \, P_S \,- \, 8 \, \left( \kappa_1^S \, + 2 \, \kappa_2^S{+2\,\kappa_3^S} \right) (\, p_1^2 \, + \, p_2^2 \, ) \, P_S \, \nonumber \\
& & -  \, 4 \, \bigg( \left( \kappa_1^{VV}+2\kappa_2^{VV}+2\kappa_3^{VV} \right) (\, p_1^2 \, + \, p_2^2 \,) \,
-  \, \left(  \kappa_1^{VV}+2\kappa_3^{VV} \right) q^2   \bigg) \, P_V(p_1^2) \, P_V(p_2^2) \, \nonumber \\
& & +  \, 4 \sqrt{2} \, \left( 2\kappa_1^{V}+2\kappa_3^{V}+\kappa_4^{V}+\kappa_5^{V}\right) \left( p_1^2 \, P_V(p_1^2) \, + \, p_2^2 \, P_V(p_2^2) \, \right) \,  \nonumber \\
& & + \, 4 \sqrt{2} \, \left( 2\kappa_3^{V}-\kappa_4^{V}+\kappa_5^{V}\right)q^2 \left(P_V(p_1^2) \, + \, P_V(p_2^2) \, \right) \,  \nonumber \\
& & - \, 4 \sqrt{2} \, \left( 2\kappa_3^{V}-\kappa_4^{V}+\kappa_5^{V}\right) \left(p_1^2\,P_V(p_2^2) \, + \,p_2^2\, P_V(p_1^2) \, \right) \,  ,
\end{eqnarray}
and
\begin{eqnarray} \label{eq:gsvv}
{\cal G}_V(p_1^2,p_2^2,q^2) & = &  8 \, \left( \kappa_1^{SVV} \, - \, 2 \, \kappa_2^{SVV} \right) \, P_S \, P_V(p_1^2) \, P_V(p_2^2) \, -\, 32 \, \kappa_2^S \, P_S \, \nonumber \\
& & + \, 8 \, \sqrt{2} \, \left( \kappa_1^{SV} \, - \, \kappa_2^{SV} \, \right) \, P_S \, \left( \, P_V(p_1^2) \, + \, P_V(p_2^2) \, \right)\, \nonumber \\
& & + \, 8 \, \sqrt{2} \, \left( \kappa_1^{V} \, - \, \kappa_2^{V} \, \right) \, \left( \, P_V(p_1^2) \, + \, P_V(p_2^2) \, \right)\,
 - \,16 \,\kappa_2^{VV} \, P_V(p_1^2) \, P_V(p_2^2) \, ,
\end{eqnarray}
where $P_S$ has been defined in Eq.~(\ref{eq:polos}) and
\begin{equation} \label{eq:polos2}
P_V(p^2) \, = \, \frac{F_V \, - \, 2 \, \sqrt{2} \, \lambda_{22}^V \, p^2}{M_V^2 \, - \, p^2} \, ,
\end{equation}
with $M_V$ the mass of the nonet of vector resonances in the $U(3)$ and chiral limit.
\par
By imposing the constraints on Eqs.~(\ref{eq:sh1v},\ref{eq:sh2v},\ref{eq:sh3v}) and (\ref{eq:sh4v}), we obtain:
\begin{eqnarray} \label{eq:svvres}
\kappa_2^S=\kappa_2^{VV} & = & 0 \, , \nonumber \\
\kappa_1^{S} \, + \, 2 \, \kappa_3^{S} \, &=& \, 0 \, , \nonumber \\
\kappa_1^{VV} \, + \, 2 \, \kappa_3^{VV} \, &=& \, 0 \, , \nonumber \\
\kappa_1^{SV} \, - \, \kappa_2^{SV} \, &=& \, 0 \, , \nonumber \\
2\,\kappa_3^{SV} \, + \, \kappa_4^{SV}\,-\,\kappa_5^{SV} \, &=& \, -\frac{2\sqrt{2}\,\lambda_{15}^S}{F_V} \, , \nonumber \\
2\,\kappa_1^{SV} \, - \,2\, \kappa_3^{SV}\, + \, \kappa_4^{SV}\, +\,\kappa_5^{SV} \, &=& \, 0 \, , \nonumber \\
 2\, \kappa_3^{SV}\, - \, \kappa_4^{SV}\, +\,\kappa_5^{SV} \, &=& \, \frac{4\,\lambda_6^V}{c_m}+\frac{M_V^2}{c_m}(2\kappa_1^V+2\kappa_3^V+\kappa_4^V+\kappa_5^V) \, ,
\nonumber \\
\kappa_1^{V} \, - \, \kappa_2^{V} \, &=& \, 0 \, , \nonumber \\
2\,\kappa_3^{V} \, - \, \kappa_4^{V}\, +\,\kappa_5^{V} \, &=& \,0 \, , \nonumber \\
\kappa_1^{V} \, + \,2 \,\kappa_3^{V} \, + \,\kappa_5^{V}&=& \, -\frac{\sqrt{2}\,\hat{C}_{61}}{F_V} \, , \nonumber \\
\kappa_1^{SVV} \, - \, 2 \, \kappa_2^{SVV} \! \! & = & \! \! \frac{F^2}{4 \, c_m \, F_V^2} \, \nonumber \\
2\,\kappa_3^{SVV} \, -\,4 \, \kappa_4^{SVV} \, + \,2\, \kappa_5^{SVV} \, + \, \kappa_6^{SVV}  & = & -\,\frac{4\,\lambda_6^{VV}}{c_m}\,+\,
\frac{F^2}{4\,c_m \, F_V^2}  \, ,
\nonumber \\
2 \, \kappa_2^{SVV} \, + \, 2 \kappa_3^{SVV} \, + \, \kappa_6^{SVV}  & = &  - \, \frac{F^2}{4 \, c_m \, F_V^2} \, - \,
\, \frac{4\sqrt{2} \lambda_3^{SV}}{F_V} \,-\,\frac{\sqrt{2}\,M_S^2\left( 2\,\kappa_3^{SV} \, - \, \kappa_4^{SV}\, + \, \kappa_5^{SV} \right)}{F_V}\nonumber \\
&  & \,-\,\frac{\sqrt{2}\,M_V^2\left( 2\,\kappa_3^{SV} \, + \, \kappa_4^{SV}\, -  \, \kappa_5^{SV} \right)}{F_V} .
\end{eqnarray}
Notice that, in this case, the local contribution from ${\cal L}_{(6)}^{\mbox{\tiny GB}}$, namely $\hat{C}_{61}$, is not forced to vanish
by the short-distance constraints. Our Lagrangian, defined in Eq.~(\ref{eq:lagsvv}),  generates both ${\cal F}_V$ and ${\cal G}_V$ functions, and is able to satisfy the short-distance relations.
\par 
Incidentally, the matching procedure in Eqs.~(\ref{eq:sh2v},\ref{eq:sh3v}) provides an expression for the vector-tensor correlator defined in Eq.~(\ref{eq:pivvdef}), namely:
\begin{equation} \label{eq:pivvres}
\Pi_{VT}(p^2) \, = \, \frac{B_0 \, F^2}{p^2 - M_V^2} \, ,
\end{equation}
that agrees with the result in \cite{Mateu:2007tr}.

\subsection{RChT coupling constants} \label{ss:rchtc}
{The relations between the RChT couplings obtained in Eqs.~(\ref{eq:saares},\ref{eq:svvres}) rely on the assumptions of short-distance QCD asymptotic behavior} and single resonance approximation. We may wonder how reliable {are those assumptions.} If our implementation of large-$N_C$ was exact ({i.e. if we had included an infinite number of resonances) we could argue that our computation should receive $\sim33 \, \%$ one-loop corrections}. In practice this is a rough estimate because we cannot evaluate the error introduced {by imposing the asymptotic behavior.} {Because of these uncertainties, one should expect slight modifications to the relations obtained in Eqs.~(\ref{eq:saares},\ref{eq:svvres}).} In our opinion {the largest source of uncertainty arises from} the lack of a more thorough implementation of the large-$N_C$ description.
\par
It is well known that the phenomenology of hadron processes indicates that large-$N_C$ is a reasonable assumption for spin-1 related processes, but fails for scalar (vacuum) quantum numbers.~\footnote{As a general setting, meson-vector form factors are well described in a $N_C  \rightarrow \infty$ framework in RChT. On the contrary, a resummation of many loops is usually required to provide a reasonable account of scalar form factors.} In this case, higher-order $1/N_C$ corrections seem to be particularly relevant. Let us consider, for instance, the case of the $c_d$ and $c_m$ couplings in Eq.~(\ref{eq:rcht2}) with the constraints in Eq.~(\ref{eq:cond2point}). One would conclude that in the single resonance approximation we have:
\begin{equation} \label{eq:cdcm}
c_d \, = \, c_m \, = \, \frac{F}{2} \, .
\end{equation}
Taking $F = 92.4 \, \mbox{MeV}$ we get $c_d=c_m = 46.2 \, \mbox{MeV}$. However, the phenomenology of different processes ($I=1/2$ and $I=3/2$ K$\pi$ s-wave scattering, $a_0(980)$ decay) gives $13 \, \mbox{MeV} \, \lsim \, c_d \, \lsim \, 40 \, \mbox{MeV}$ and $30 \, \mbox{MeV} \, \lsim \, c_m \, \lsim \,  100 \, \mbox{MeV}$ (see~\cite{Escribano:2010wt} and references therein). {While the condition $4 \, c_d \, c_m \, = \, F^2$ is rather well satisfied, there seems to be some tension between the phenomenological values of $c_d$ and $c_m$ and the relation $c_d=c_m$. Given the large uncertainties, we cannot reliably estimate the error of our large-$N_C$ result (\ref{eq:cdcm}) (in single resonance approximation), but it could be off even by a factor of 3 (for $c_d$) or 2 (for $c_m$) in the worst case.}
\par
We conclude that our relations in Eq.~(\ref{eq:saares},\ref{eq:svvres}) may be affected by errors of similar size to the case above. The order of magnitude is expected to be correct but notable deviations may arise. Unfortunately, we cannot constrain most of the couplings with the present phenomenological status. However we can get reliable estimates in certain couplings, such as $c_d$ and $c_m$, which appear in the decays of a scalar to two pseudoscalars. We will pursue this in Section~\ref{s:scalars}.
\par
In summary, our present knowledge of the hadron scalar spectrum, and its decays, is rather poor \cite{Tanabashi:2018oca} and the couplings involved are essentially unknown. On one hand, we need to identify which is the spectrum described by the RChT (or any other) framework. On the other hand, we lack the required experimental data to have a general vision of the accuracy of our results. In the next section we will try to clarify part of the phenomenological status of scalar resonances.

\section{Scalar couplings} \label{s:scalars}
Which are the, experimentally identified, scalar states present in our Lagrangian? As commented at the end of Section~\ref{s:rcht}, there is almost no discussion on the identification of the vector and axial-vector resonances of the RChT Lagrangian. They are, in fact, the lightest hadron resonances in the spectrum with those quantum numbers. Scalars (and glueballs) are different. They carry the vacuum quantum numbers and their identification (for $M \lsim \, 2 \, \mbox{GeV}$) generates controversy. Here, we will comment first several, more or less agreed,
features and we will propose a scheme.
\par
As discussed in Section~\ref{s:rcht} the lightest scalar resonance, namely the isosinglet $f_0(500)$, corresponds to a wide $\pi \pi$ s-wave that does not survive the $N_C \rightarrow \infty$ limit. Increasing in mass we have $K_0^*(700)$, the isotriplet $a_0(980)$ and the isosinglet $f_0(980)$. The next scalar appears at around $1.4 \, \mbox{GeV}$. Hence, naively, one could consider that the first $U(3)$ nonet of scalar resonances is the one with those states: $S_L \, = \, \{f_0(500), K_0^*(700), a_0(980), f_0(980) \}$. Following this scheme, determined by the mass, the next nonet would be: $S_H \, = \, \{f_0(1370),K_0^*(1430),a_0(1450),f_0(1500) \} $.
Until $ \sim 2 \, \mbox{GeV}$ there is another isosinglet scalar: $f_0(1710)$. Other scalars appear around $2 \, \mbox{GeV}$. Needless to say that the physical states do not need to correspond exactly with the basis in the Lagrangian and mixing between those with the same quantum numbers surely arise. If our assumption, relying on the mass, was correct, we could conclude that $S_L$ would correspond to the nonet that vanishes at $N_C \rightarrow \infty$, as it includes {the} $f_0(500)$. A thorough analysis in this limit was carried out in Ref.~\cite{Cirigliano:2003yq}. Their conclusion was that the most favored candidates for the leading nonet in the infinite number of colors limit was: $S_{\infty} \, = \, \{ f_0(980),K_0^*(1430),a_0(1450),f_0(1500)\}$.
\par
Another aspect of the spectrum of scalars is related with their quark content. This is of no relevance for the RChT Lagrangian: it can allocate any quark content. However it is suitable to collect this information here. We will reduce our comment to $[\bar{q}\, q]$ and $[\bar{q} \, q][\bar{q} \, q]$ states (see \cite{Jaffe:2004ph} and references therein). One aspect that distinguishes the quark structure of the nonets is that, in the ideal mixing case, the tetraquark multiplet has an inverted spectrum: the isodoublet is heavier than the isotriplet. We see that this feature {(the order in the spectrum)} is clearly described by $S_L$ above, while they are essentially degenerated (within errors \cite{Tanabashi:2018oca}) in the case of $S_H$. This feature could be the result of a violation of the ideal mixing. There are also other reasons to conclude that the light nonet corresponds to the tetraquark structure while
the heavy one is the usual $[\bar{q} q]$~\cite{Jaffe:2004ph}.
\par
In this section we will identify the nonet of scalar resonances in our RChT Lagrangian with the $S_H$ nonet above. We will also consider the singlet $f_0(1710)$ and a general mixing between the isosinglet fields that generates the physical states, including a possible glueball.  As commented in Subsection~\ref{ss:rchtc}, the phenomenology seems to indicate that the $N_C \rightarrow \infty$ limit is rather poor when scalars are involved. Hence, in our analysis, we will include subleading contributions into the Lagrangian in order to accommodate the experimental figures within their large errors. This will allow {us} to get more accurate determinations of the leading $c_d$ and $c_m$ couplings.
\par
Similar studies have been carried out in the last years, see for instance~\cite{Black:1998wt,Black:1999yz,Chen:2009zzs,Zhou:2010ra,Fariborz:2015dou,Fariborz:2015era,Noshad:2018afw,Kim:2018zob} and references therein.

\subsection{$\mathbold{S \rightarrow PP}$: isodoublet and isotriplet decays} \label{ss:isodtrip}
We will consider a RChT framework with violation of the $N_C \rightarrow \infty$ limit in the tree level Lagrangian. More precisely, we will consider
terms with more than one trace in flavour space. Previous studies \cite{Black:1999yz} have pointed out a non-negligible mixing between the $I=1,1/2$ states of both nonets $S_L$ and $S_H$. Hence we will include a mixing between them. The Lagrangian reads:
\begin{eqnarray} \label{eq:lagdt}
{\cal L}_{I=1,1/2} \, & = &  \, c_d^L \, \langle \, S_L \, u_{\mu} \, u^{\mu} \, \rangle \, + \, \alpha_L \, \langle \, S_L \, u_{\mu} \, \rangle
\, \langle u^{\mu} \, \rangle \, + \, c_m^L \, \langle \, S_L \, \chi_{+} \, \rangle  \nonumber \\
& &  \; + \, c_d^H \, \langle \, S_H \, u_{\mu} \, u^{\mu} \, \rangle \, + \, \alpha_H \, \langle \, S_H \, u_{\mu} \, \rangle \, \langle \,
u^{\mu} \, \rangle \, + \, c_m^H \, \langle \, S_H \, \chi_{+} \, \rangle \, ,
\end{eqnarray}
after diagonalization. This introduces two mixing angles:
\begin{eqnarray} \label{eq:2nonrot}
\left( \begin{array}{c}
          a_{0,L} \\
          a_{0,H}
          \end{array}  \right) & = & \left( \begin{array}{cc}
                                   \cos \varphi_a & \sin \varphi_a \\
                                   - \sin \varphi_a & \cos \varphi_a
                                   \end{array} \right) \;
                                   \left( \begin{array}{c}
                                              a_0(980) \\
                                              a_0(1450)
                                              \end{array} \right)  \, \nonumber  ,  \\
                                              & & \\
\left( \begin{array}{c}
          K^*_{0,L} \\
          K^*_{0,H}
          \end{array} \right)  & = & \left( \begin{array}{cc}
                                   \cos \varphi_k & \sin \varphi_k \\
                                   - \sin \varphi_k & \cos \varphi_k
                                   \end{array} \right) \;
                                   \left( \begin{array}{c}
                                              K_0^*(700) \\
                                              K^*_0(1430)
                                              \end{array} \right)  \,   . \nonumber
\end{eqnarray}
The mixing angles $\varphi_a$ and $\varphi_k$ are not fixed. In Ref.~\cite{Black:1999yz} the values quoted are $\varphi_a = \pi/4$ and
$\varphi_k \sim 0.17 \, \pi$. We will consider them as free parameters. The lack of data on the FSI phase shifts for {the decays of these fields} prevents the inclusion of these effects in our analysis. The amplitudes for such decays are collected in Subsection \ref{ssap:triplet} of Appendix~\ref{app:damp}.

\subsection{$\mathbold{S \rightarrow PP}$: isosinglet decays} \label{ss:isos}
As commented before, we are interested in the description of the decays of the $f_0(1370)$, $f_0(1510)$ and $f_0(1710)$. Although we identify the {first two} as those of the $S_H$ multiplet and the third as a possible glueball, the real situation can be much more cumbersome and the real physical states is surely a non-neglible mixing between the isosinglets of the $S_H$ multiplet (namely $S_8$,$S_0$) and an extra singlet ($S_1$). A general rotation of them will provide the physical states:
\begin{equation}
\label{eq:diag}
\left( \begin{array}{c} f_0(1370) \\ f_0(1510) \\ f_0(1710) \end{array} \right) \; = \; A \; \left( \begin{array}{c} S_8 \\ S_0 \\ S_1 \end{array} \right) ,
\end{equation}
where
\begin{equation}
\label{eq:matrixa}
A \; =  \; \left( \begin{array}{ccc}
            \cos \gamma \cos \beta \cos \alpha - \sin \gamma \sin \alpha & \cos \gamma \cos \beta \sin \alpha + \sin \gamma \cos \alpha & - \cos \gamma \sin \beta
            \\
            -\sin \gamma \cos \beta \cos \alpha - \cos \gamma \sin \alpha & - \sin \gamma \cos \beta \sin \alpha + \cos \gamma \cos \alpha & \sin \gamma \sin \beta
            \\
            \sin \beta \cos \alpha & \sin \beta \sin \alpha & \cos \beta
            \end{array} \right).
\end{equation}
{Now we} set up our RChT framework to describe these decays. Contrary to the first decays, we are not going to consider mixing between the light and heavy multiplets. This would give a complicated setting with many parameters and, as we will conclude, it is not necessary to provide a reasonable description of {all} the decays.
\par
With these inputs the Lagrangian to study the $f_0 \rightarrow P P$ decays will be:
\begin{eqnarray}
\label{eq:intu5}
{\cal L}_{I=0,S_1} &=& c_d^H \, \langle \, S_H \, u_{\mu} \, u^{\mu} \, \rangle \, + \, c_{m}^H \, \langle \,  S_H \, \chi_+\,  \rangle\,
+ \,  \alpha_H \, \langle \, S_H \,  u_{\mu} \, \rangle \, \langle \,  u^{\mu} \,  \rangle \, + \,  \beta_H \, \langle \, S_H \, \rangle \, \langle \, u_{\mu} \, u^{\mu} \, \rangle \, \nonumber\\
&&  + \, \gamma_H \, \langle \, S_H \, \rangle \, \langle \,  u_{\mu} \,  \rangle \, \langle \, u^{\mu} \, \rangle \,
+ \, c_d' \, S_1 \, \langle \, u_{\mu} \, u^{\mu} \, \rangle \, + \, c_m' \, S_1 \, \langle \, \chi_+ \,  \rangle \, + \, {\gamma'} \, S_1 \, \langle \, u_{\mu} \,  \rangle \,  \langle \,  u^{\mu} \,  \rangle \, .
\end{eqnarray}
Furthermore, as the $\pi \pi$ and $\overline{K}K$ phase shifts are rather well known \cite{Dai:2014lza,Dai:2014zta} we also incorporate the
parameterization of final state interactions as described in Appendix~\ref{app:fsi}. The amplitudes for these decays are gathered in Subsection~\ref{ssap:singlet} of Appendix~\ref{app:damp}.

\subsection{Results} \label{ss:results}
The present experimental determination of the $S \rightarrow PP$ decay widths is rather poor. Many channels have not been observed or have large errors. 
As a result, we end up with more variables than experimental inputs. However, from our fit we can obtain a general idea of the current landscape.
\par
We will fit our partial widths and ratios with the data collected in the rightmost column of Tables~\ref{tab:widths} and \ref{tab:ratios}.
\begin{table}
\begin{center}
\begin{tabular}  {|c|c|c|}
\hline
Width           &                      Our fit (MeV)  &  Exp. (MeV)                 \\
\hline
\hline
$\Gamma_{f_0(1370)\to \pi\pi}$       & $11.7\pm5.7$  &  $20.8\pm10.7$ \cite{Bugg:1996ki,Bargiotti:2003ev} \\
$\Gamma_{f_0(1370)\to K \bar{K}}$   & $10.7\pm3.2$  &  $19.0\pm10.6$ \cite{Bugg:1996ki,Bargiotti:2003ev} \\
$\Gamma_{f_0(1370)\to\eta\eta}$     & $10.4\pm4.3$  &  $6.41\pm2.88$  \cite{Abele:2001pv,Abele:2001js}\\
\hline
\hline
$\Gamma_{f_0(1500)\to \pi\pi}$         & $38.1\pm5.6$  &  $38.0\pm2.5$    \cite{Albaladejo:2008qa,Ablikim:2006db}    \\
$\Gamma_{f_0(1500)\to K \bar{K}}$      & $9.39\pm2.2$  &  $9.37\pm1.09$ \cite{Albaladejo:2008qa,Ablikim:2006db}       \\
$\Gamma_{f_0(1500)\to\eta\eta}$        & $5.50\pm4.1$  &  $5.56\pm0.98$  \cite{Albaladejo:2008qa,Ablikim:2006db} \\
$\Gamma_{f_0(1500)\to\eta\eta'}$       & $0.0$   &  $2.07\pm0.87$  \\
\hline
\hline
$\Gamma_{f_0(1710)\to \pi\pi}$       & $20.5\pm6.6$  &  $20.5\pm9.9$       \\
$\Gamma_{f_0(1710)\to K \bar{K}}$     & $50.0\pm15.3$  &  $50.0\pm16.7$      \\
$\Gamma_{f_0(1710)\to\eta\eta}$      & $23.8\pm9.8$  &  $24.0\pm11.0$  \\
$\Gamma_{f_0(1710)\to\eta\eta'}$      & $30.9\pm20.2$  &  $-$  \\
\hline
\hline
$\Gamma_{a_0^+(1450)\to\pi^+ \eta}$    & $24.4\pm12.0$  &  $24.7\pm5.3$  \\
$\Gamma_{a_0(1450)\to\pi^0 \eta}$     & $24.5\pm12.0$  & $24.7\pm5.3$  \\
$\Gamma_{a_0^+(1450)\to\pi^+ \eta'}$   & $9.14\pm7.6$  &  $8.7\pm4.5$  \\
$\Gamma_{a_0(1450)\to\pi^0 \eta'}$    & $9.18\pm7.7$  &  $8.7\pm4.5$  \\
$\Gamma_{a_0^+(1450)\to K^+\overline{K}^0}$  & $21.0\pm7.3$  &  $21.7\pm7.4$  \\
$\Gamma_{a_0^0(1450)\to K^+K^-}$      & $10.6\pm3.7$  &  $-$  \\
$\Gamma_{a_0^0(1450)\to K^0\overline{K}^0}$  & $10.4\pm3.6$  &  $-$  \\
\hline
\hline
$\Gamma_{{K_0^*}^+(1430)\to\pi^0 K^+}$   & $80.5\pm12.8$  & $-$  \\
$\Gamma_{{K_0^*}^+(1430)\to\pi^+K^0}$  & $159.7\pm25.5$ & $-$  \\
$\Gamma_{{K_0^*}^0(1430)\to\pi^0 K^0}$  & $80.0\pm12.8$  &  $-$  \\
$\Gamma_{{K_0^*}^0(1430)\to\pi^-K^+}$  & $160.6\pm25.6$ &  $-$  \\
$\Gamma_{{K_0^*}^+(1430)\to\eta K^+}$   & $20.7\pm14.3$  &  $-$  \\
$\Gamma_{{K_0^*}^0(1430)\to\eta K^0}$   & $20.5\pm14.2$  &  $-$  \\
$\Gamma_{{K_0^*}^+(1430)\to\pi K}$    & $240.1\pm38.3$ &  $251.1\pm27.0$  \\
\hline
\hline
$\Gamma_{a_0^+(980)\to\pi^+ \eta}$    & $81.2\pm16.9$  &  $-$  \\
$\Gamma_{a_0(980)\to\pi^0 \eta}$       & $81.7\pm17.0$  &  $-$  \\
$\Gamma_{a_0^+(980)\to K^+\overline{K}^0}$ & $14.4\pm5.5$  &  $14.2\pm1.8$  \\
$\Gamma_{a_0^0(980)\to K^+K^-}$        & $7.66\pm2.8$  &  $-$  \\
$\Gamma_{a_0^0(980)\to K^0\overline{K}^0}$   & $6.68\pm2.7$  &  $-$  \\
\hline
\hline
$\Gamma_{{K_0^*}^+(700)\to\pi^0 K^+}$ & $1.56\pm1.9$  &  $-$  \\
$\Gamma_{{K_0^*}^+(700)\to\pi^+K^0}$   & $3.04\pm3.6$  &  $-$  \\
$\Gamma_{{K_0^*}^0(700)\to\pi^0 K^0}$       & $1.53\pm1.8$  &  $-$  \\
$\Gamma_{{K_0^*}^0(700)\to\pi^-K^+}$         & $3.09\pm3.7$  &  $-$  \\
$\Gamma_{{K_0^*}(700)\to\pi K}$    & $4.59\pm5.5$  & $478\pm127$  \\
\hline
\hline
 \end{tabular}
 \caption{\label{tab:widths} Results of our fit for the decay widths analysed in our RChT framework. The experimental data are taken from \cite{Tanabashi:2018oca} except when explicitly stated otherwise.}
 \end{center}
\end{table}
\begin{table}
\begin{center}
\begin{tabular}  {|c|c|c|c|}
\hline
Decaying particle                & Ratio   &  Our fit  &  Exp.                   \\
\hline
\hline
$f_0(1370)$ & $\mbox{Br}[K \overline{K} / \pi\pi]$          &  $0.912\pm0.374$ &  $0.91\pm0.20$  \cite{Bargiotti:2003ev} \\
& $\mbox{Br}[\eta\eta / \pi\pi]$                & $0.889\pm0.771$ &  $0.31\pm0.80$ \cite{Abele:2001pv,Abele:2001js} \\
\hline
\hline
$f_0(1500)$ & $\mbox{Br}[K \overline{K} / \pi\pi]$        & $0.246\pm0.006$ &  $0.246\pm0.026$  \\
& $\mbox{Br}[\eta\eta / \pi\pi]$                   & $0.144\pm0.002$ &  $0.145\pm0.027$  \\
& $\mbox{Br}[\eta'\eta / \pi\pi]$                  & $0.0$   & $0.055\pm0.024$  \\
\hline
\hline
$f_0(1710)$ &  $\mbox{Br}[\pi\pi/K \overline{K} ]$            & $0.410\pm0.037$ &  $0.41\pm0.14$ \cite{Albaladejo:2008qa,Ablikim:2006db}  \\
& $\mbox{Br}[\eta\eta / K \overline{K}]$           & $0.476\pm0.282$ &  $0.48\pm0.15$  \\
\hline
\hline
$a_0(1450)$ & $\mbox{Br}[\pi \eta' / \pi \eta]$                & $0.375\pm0.163$& $0.35\pm0.16$ \\
& $\mbox{Br}[K \overline{K} / \pi \eta]$             & $0.859\pm0.269$ &  $0.88\pm0.23$ \cite{Abele:2001pv} \\
\hline
\hline
$K_0^*(1430)$ &$\mbox{Br}[ \eta K /\pi K ]$                   & $0.086\pm0.074$ &  $0.092\pm0.031$ \cite{Lees:2014iua} \\
\hline
\hline
$a_0(980)$ & $\mbox{Br}[K \overline{K} / \pi \eta]$           & $0.175\pm0.057$ &  $0.183\pm0.024$  \\
\hline
\hline
 \end{tabular}
 \caption{\label{tab:ratios} Results of our fit for the ratios of decay widths analysed in our RChT framework. The experimental data are taken from \cite{Tanabashi:2018oca} except when explicitly stated otherwise.}
\end{center}
\end{table}
We input the masses of the resonances from \cite{Tanabashi:2018oca}, with the exception of the $a_0(980)$ and $f_0(1370)$. The first one is also fitted due to the sensibility of the results to its decay. For $f_0(1370)$ we take the result put forward by \cite{Abele:2001js} in the analysis of its dominant decay into four pions, $M_{f_0(1370)} = 1.395 \, \mbox{GeV}$. We take $F = 92.4 \, \mbox{MeV}$ for the decay constant of the pion.
\par
Our results for the fit are presented in the central column of Tables~\ref{tab:widths} and \ref{tab:ratios}. As we can see, we obtain a reasonable description of most of the channels (being the clear exception the $K_0^*(700) \rightarrow \pi K$ decay). We get a null value
for $\Gamma(f_0(1500) \rightarrow \eta \eta')$ since this decay is kinematically forbidden for the central value of the $f_0(1500)$ mass.
The results for masses, couplings and parameters are collected in Table~\ref{tab:const;fit;all}.
\begin{table}
\begin{center}
\begin{tabular}  {|c|c|c|c|}\hline
Parameter                   &  Our fit  & Mixing angle                   &  Our fit  \\
\hline
\hline
$M_{a_0(980)}$     &  $1023.8\pm22.6$    &  &  \\
\cline{1-2}
$c_d^L$               & $15.6\pm1.9$  &  $\alpha$    & $-98.8\pm41.9$        \\
$c_d^H$             & $3.07\pm1.00$  & $\beta$       & $-39.8\pm13.7$   \\
$c_d'$             & $0.0$          &  $\gamma$      & $-27.8\pm44.4$   \\
$c_m^L$             & $13.3\pm6.8$   & $\omega$      & $53.6\pm4.7$   \\
$c_m^H$           & $9.21\pm3.21$   & $\varphi_a$    & $4.78\pm3.75$ \\
$c_m'$           & $0.0$         &  $\varphi_k$    & $90.3\pm22.5$  \\
$\alpha_L$        & $17.9\pm3.2$  &  &    \\
$\alpha_H$           & $0.88\pm1.50$   &  &   \\
$\beta_H$        & $-3.42\pm0.53$     &  &     \\
$\gamma_H$           & $-6.45\pm1.19$    &  &      \\
$\gamma'$       & $1.43\pm3.26$    &  &      \\
\hline
\hline
$\chi^2_{d.o.f}$          & $0.40$     &  &      \\
\hline
\hline
 \end{tabular}
 \caption{\label{tab:const;fit;all} Results of the fit for the parameters in the RChT framework. The mass and all the couplings are given in MeV. All the angles are in degrees.}
\end{center}
\end{table}
We are going to analyse, in turn, the outcome:
\begin{itemize}
\item[a)] We obtain the mixing angles between the $I=0$ states, $\alpha$, $\beta$, $\gamma$ with rather large errors. To illustrate the results let us change to the flavour basis, $|S\rangle$, $|N\rangle$, $|G\rangle$, defined by:
\begin{eqnarray} \label{eq:sng}
| S \rangle & \equiv & | \overline{s}s\rangle \, = \, - \sqrt{\frac{2}{3}} \, | S_8 \rangle + \frac{1}{\sqrt{3}} \, | S_0  \rangle \, , \nonumber \\
| N \rangle & \equiv & \frac{1}{\sqrt{2}}| \overline{u}u + \overline{d}d \rangle \, = \,  \frac{1}{\sqrt{3}} \,  | S_8 \rangle + \sqrt{\frac{2}{3}} \, | S_0  \rangle ,
\end{eqnarray}
being $|G\rangle$ the singlet glueball. In this basis {we have}:
\begin{equation} \label{eq:mixinghn2}
\left( \begin{array}{c}
f_0(1370) \\
f_0(1500) \\
f_0(1710)
\end{array} \right)
\; = \,
\left( \begin{array}{ccc}
-0.82 \pm 0.22 & 0.12 \pm 0.49 & 0.57\pm 0.16 \\
0.07 \pm 0.48 & -0.95 \pm 0.24 & 0.30 \pm 0.25 \\
0.57 \pm 0.14 & 0.29 \pm 0.23  & 0.77 \pm 0.09
\end{array} \right)\;
\left( \begin{array}{c}
N \\
S \\
G \\
\end{array}
\right) .
\end{equation}
From this result we conclude that there is a dominant one-to-one identification between $f_0(1370)$,$f_0(1500)$ and $f_0(1710)$ with $N$, $S$ and $G$ respectively. Notwithstanding there seems to be also a large mixing between $f_0(1370)$ and $f_0(1710)$ with the $N$ and $G$ states.
\par
Our result agrees with solution II of Ref.~\cite{Giacosa:2005zt}. Their solution I switches the roles of $f_0(1500)$ and $f_0(1710)$. Different models and different settings can be found in the literature. Our conclusion differs from the one in Ref.~\cite{Chen:2009zzs} because although they agree on identifying the $f_0(1710)$ mostly with the glueball, they find that $f_0(1370)$ is dominantly $| S \rangle$ and $f_0(1500)$ is dominantly $|N\rangle$. This later identification of $f_0(1370)$ is also found in Ref.~\cite{Fariborz:2015era}, though with a noticeable four-quark component too. In Ref.~\cite{Fariborz:2015dou} {it} was concluded that $f_0(1500)$ was mostly glueball but $f_0(1710)$ was also sharing a large component. Ref.~\cite{Chatzis:2011qz} provides two scenarios: In one of them $f_0(1710)$ is dominantly glueball; in the other this role corresponds to $f_0(1500)$.
\par
In relation with the mixing between the light and heavy nonets of scalar resonances, our results differ from those of Ref.~\cite{Black:1999yz},  and we find a tiny mixing for the $a_0$ states and an almost inverted situation for the $K_0^*$ states.
\item[2/] The couplings in Eq.~(\ref{eq:intu5}), $c_d'$,$c_m'$ and $\gamma'$, involving the extra singlet $S_1$ (glueball), are consistent with zero. This indicates that the glueball component only arises through the mixing with the $I=0$ singlets of the nonet.
\item[3/] The rest of RChT couplings show an interesting trend. Although with large errors, {the} expected  $1/N_C$ suppression  between the leading and next-to-leading terms {does not seem to be realized}. They are essentially of the same order. We verify that both multiplets satisfy the condition in Eq.~\eqref{eq:cond2point}: $c_d^L \, c_m^L > 0$ and $c_d^H \, c_m^H > 0$, but we notice that the relation $c_d=c_m$ is approximately satisfied only by the light multiplet $c_d^L \sim c_m^L$. Meanwhile the heavy multiplet deviates from this relation. {None} of them satisfies, numerically, Eq.~(\ref{eq:cdcm}), though the light multiplet comes close.
\end{itemize}

\section{Conclusions} \label{s:conclud}
The phenomenology of the lightest hadron scalars is rather clumsy. The issues of identification of the $U(3)$ nonets, its nature and their decays embrace a thorough research and a large number of publications. Many aspects remain to be understood. In this work we have tried to put some light on the features and problems that have to be taken into account for a Lagrangian description of the scalar sector; in our case  within the Resonance Chiral Theory.
\par
The greater part of the decays of scalar resonances involve the $\langle \, S \, V_{\mu} \, V_{\nu} \, \rangle$ and $\langle \, S \, A_{\mu} \,A_{\nu} \, \rangle$ Green functions of QCD currents. We have analysed these within RChT, including the necessary operators in order to fulfill the {short-distance requirements determined by the matching} in Eq.~(\ref{eq:match}). As a result we found a set of relations between the couplings in our Lagrangian. These should be valid in the $N_C \rightarrow \infty$ limit and single resonance approximation.
Although the procedure that we have followed has given in the past many  successful predictions, we know that hadron scalar-involved amplitudes are not well behaved in the large-$N_C$ limit. {In order to assess our results, we have carried out a fit to $S \rightarrow PP$ decays in Section~\ref{s:scalars}. In the fit we have included subleading contributions in $1/N_C$, to analyze the behavior of our RChT description of such decays.} The results of our study are indeed pointing out that operators that should be suppressed following large-$N_C$ premises are in fact as relevant as the leading ones. Hence, at least part of the relations between the couplings involving scalars, in the $N_C \rightarrow \infty$ limit, may be largely violated. We have to stress, though, that the poor, and sometimes confusing, experimental determinations in most of the scalar decays could mislead this conclusion. It will be important to improve the experimental measurements in order to validate this scenario.
\par
As a consequence of our study we also conclude that, within errors, $f_0(1370)$ is dominantly a
$|\overline{u}u + \overline{d}d\rangle$ state, $f_0(1710)$ is dominantly a glueball, but both of them also have a noticeable mixing.
{The $f_0(1500)$ is dominantly a $|\overline{s}s\rangle$ state.}
The results by other authors vary, {however the use of different frameworks make the comparison difficult.}
\par
The study of hadron scalar resonances remains an open field. Their spectrum, classification and nature originate a rich debate. The large-$N_C$ framework, already questioned in the study of these decays, does not seem to be the proper setting because of the large size of subleading corrections. However a solid conclusion will only be possible if a better experimental knowledge of the spectrum and decays is achieved.

\section*{Acknowledgements}
We wish to thank Gerhard Ecker and Roland Kaiser for their participation in an early stage of this project.
This work has been supported in part by Grants No. FPA2014-53631-C2-1-P, FPA2017-84445-P and SEV-2014-0398 (AEI/ERDF, EU) and by PROMETEO/2017/053 (GV).
Ling-Yun Dai thanks the support from National Natural Science Foundation of China (NSFC) with Grant No. 11805059 and the Fundamental Research Funds for the Central Universities with Grant No. 531107051122. The work of J.F. was supported in part by the Swiss National Science Foundation (SNF) under contract 200021-159720.

\appendix
\renewcommand{\theequation}{\Alph{section}.\arabic{equation}}
\renewcommand{\thetable}{\Alph{section}.\arabic{table}}
\vspace{0.7cm}
\noindent
\section*{Appendices}
\setcounter{equation}{0}
\setcounter{table}{0}
\section{Chiral notation} \label{app:rcht}
We collect briefly the basic notation used in both ChPT and RChT \cite{Cirigliano:2006hb}. The Goldstone fields $\phi$ parameterize the elements $u(\phi)$ of the
coset space $SU(3)_L \otimes SU(3)_R / SU(3)_V$: 
\begin{equation} \label{eq:a_u}
u(\phi) = \exp{\left\{\frac{i}{\sqrt{2} F} \, \Phi(\phi) \, \right\}} \ ,
\end{equation}
where $F$ is the decay constant of the pion in the chiral limit and
\begin{equation} \label{eq:a_phi}
\Phi (\phi) \, = \,   \sum_{i=1}^8 \, \lambda_i \, \frac{\phi_i}{\sqrt{2}} \, = \,
\left(
\begin{array}{ccc}
 \displaystyle\frac{1}{\sqrt 2}\,\pi^0 + \displaystyle\frac{1}{\sqrt
 6}\,\eta_8
& \pi^+ & K^+ \\
\pi^- & - \displaystyle\frac{1}{\sqrt 2}\,\pi^0 +
\displaystyle\frac{1}{\sqrt 6}\,\eta_8
& K^0 \\
 K^- & \bar{K}^0 & - \displaystyle\frac{2}{\sqrt 6}\,\eta_8
\end{array}
\right)
\ ,
\end{equation}
with $\lambda_i$ the Gell-Mann matrices.
\par
The nonlinear realization of $SU(3)_L \otimes SU(3)_R$ on resonance fields depends on their transformation  properties
under the unbroken $SU(3)_V$, the flavour group. Here we will consider massive states transforming as octets ($R_8$) or singlets ($R_0$),
with $R=V,A,S,P$ for vector, axial-vector, scalar and pseudoscalar fields, respectively. In the large-$N_C$ limit both become degenerate in the chiral limit and we collect them in a nonet field:
\begin{equation} \label{eq:a_resona}
R =  \sum_{i=1}^{8} \, \lambda_i \, \frac{R_i}{\sqrt{2}} \,  + \, \frac{R_0}{\sqrt{3}} ~\mathbbm{1}  \, .
\end{equation}
We will use the antisymmetric representation for the spin-1 fields \cite{Kyriakopoulos:1969zm}.
In order to calculate Green functions of vector, axial-vector and scalar currents, it is
convenient to include external hermitian sources $\ell_{\mu}(x)$ (left), $r_{\mu}(x)$ (right), $s(x)$ (scalar) and $p(x)$ (pseudoscalar).
\par
With the fundamental building blocks $u(\phi)$, $V_{\mu \nu}$, $A_{\mu \nu}$, $S$, $\ell_{\mu}$, $r_{\mu}$, $s$ and $p$, the hadronic Lagrangian
is given by the most general set of monomials invariant under Lorentz, chiral, P and C transformations. At leading order in $1/N_C$, the
monomials should be constructed by taking a single {trace} of products of chiral operators (exceptions to this rule are not of interest for our research). The chiral tensors $\chi(p^n)$, i.e. those not including resonance fields, can be labeled according to the chiral power counting. The independent building blocks of lowest dimension are:
\begin{eqnarray} \label{eq:a_ingr}
u_\mu &=& i \{ u^\dagger(\partial_\mu - i r_\mu)u -
u(\partial_\mu - i \ell_\mu) u^\dagger\}
\qquad \qquad \qquad  [{\cal O}(p)] \; ,
\nonumber \\
\chi_\pm &=&  u^\dagger \chi u^\dagger \pm u \chi^\dagger u
\qquad \qquad \qquad \qquad \qquad \qquad \qquad \,\,
  [{\cal O}(p^2)] \; ,
\nonumber \\
f_\pm^{\mu\nu} &=& u F_L^{\mu\nu} u^\dagger \pm u^\dagger F_R^{\mu\nu} u
\qquad \qquad \qquad \qquad \qquad \qquad    \, \, \, \,
 [{\cal O}(p^2)]  \; ,
\nonumber \\
h_{\mu\nu} &=& \nabla_\mu u_\nu + \nabla_\nu u_\mu
\qquad  \qquad \qquad \qquad \qquad \qquad \qquad  \, \,   [{\cal O}(p^2)]  \; ,
\end{eqnarray}
with
\begin{eqnarray} \label{eq:a_chi}
\chi= 2 B_0 (s+ip) \; , & & \qquad \qquad  B_0 = - \frac{\langle 0 | \overline{u} u | 0 \rangle}{F^2} \, ,
\end{eqnarray}
and non-Abelian field strengths
$F_R^{\mu\nu} = \partial^\mu r^\nu - \partial^\nu r^\mu -
i[r^\mu,r^\nu] $, $F_L^{\mu\nu} = \partial^\mu \ell^\nu - \partial^\nu
\ell^\mu - i [\ell^\mu,\ell^\nu]$. The covariant derivative is
defined by $\nabla_\mu X = \partial_\mu X + [\Gamma_\mu,X] $,
in terms of the chiral connection
$\Gamma_\mu = \{ u^\dagger (\partial_\mu - i r_\mu)u +
u (\partial_\mu - i \ell_\mu) u^\dagger \}/2$
for any operator $X$ transforming as an octet of $SU(3)_V$.
Higher-order chiral tensors can be obtained by taking products of
lower-dimensional building blocks or by acting on them with the
covariant derivative.

\setcounter{equation}{0}
\setcounter{table}{0}
\section{$\mathbold{S \rightarrow PP}$ decay amplitudes} \label{app:damp}
The widths of the $S \rightarrow P_1 P_2$ decays are given by:
\begin{equation} \label{eq:fppn8}
\Gamma (S \rightarrow P_1P_2)  \equiv \Gamma_i \; = \; \frac{\lambda^{1/2}(m_{S}^2, m_{P_1}^2, m_{P_2}^2 )}{16 N_{P_1P_2} \pi M_S^3} \,  \mid\mathcal{M}_{S\to P_1P_2}\mid^2 \, \; ,
\end{equation}
with $\lambda(a,b,c) = (a+b-c)^2-4ab$. Notice that $N_{P_1P_2}$ is 2 for two identical particles such as $\pi^0\pi^0,\eta\eta$. Here
we have taken into consideration the effect of mass of the final mesons in the phase space.
In Eq.~(\ref{eq:fppn8}) the amplitudes $\mathcal{M}_{SP_1P_2}$ are given in the following subsections.

\subsection{\B{ I = 1, 1/2} decays} \label{ssap:triplet}
The couplings and mixing for the decays of $a_0(980)$, $a_0(1450)$, $K_0^*(700)$ and $K_0^*(1430)$, have been defined in
Eqs.~(\ref{eq:lagdt},\ref{eq:2nonrot}). The decay amplitudes, defined in Eq.~(\ref{eq:fppn8}),
of the isovectors and isodoublets in the $S_H$ multiplet are:
\begin{eqnarray}
\mathcal{M}_{a_0^+\to  \eta\pi^+}^H & = &-\frac{1}{\sqrt{3} F^2} \bigg\{\, \left(M_{a_0}^2-m_\eta^2-m_{\pi^+}^2 \right) \bigg[\sqrt{2}\cos\theta(c_{d}^L\sin \varphi_a+c_{d}^H\cos\varphi_a) \nonumber \\
& & \qquad \qquad \qquad \qquad \qquad \qquad \; \; -\sin\theta \left((3\alpha_L+2c_{d}^L)\sin\varphi_a
 +(3\alpha_H+2c_{d}^H)\cos\varphi_a\right)\,\bigg] \nonumber \\
& & \qquad \qquad  + \, 2 \, m_\pi^2(\sqrt{2}\cos\theta-2\sin\theta)(c_{m}^L\sin\varphi_a+c_{m}^H\cos\varphi_a)\,\bigg\} \,,\nonumber
\\[3mm]
\mathcal{M}_{a_0^0\to  \eta\pi^0}^H & = & -\frac{1}{\sqrt{3} F^2} \bigg\{\, \left(M_{a_0}^2-m_\eta^2-m_{\pi^0}^2 \right) \bigg[\sqrt{2}\cos\theta(c_{d}^L\sin \varphi_a+c_{d}^H\cos\varphi_a) \nonumber \\
& & \qquad \qquad \qquad \qquad \qquad \qquad \; \; -\sin\theta \left((3\alpha_L+2c_{d}^L)\sin\varphi_a
+(3\alpha_H+2c_{d}^H)\cos\varphi_a\right)\,\bigg]  \nonumber \\
& & \qquad \qquad + \,2 \, m_\pi^2(\sqrt{2}\cos\theta-2\sin\theta)(c_{m}^L\sin\varphi_a+c_{m}^H\cos\varphi_a)\,\bigg\} \,, \nonumber
\\[3mm]
\mathcal{M}_{a_0^+\to  \eta'\pi^+}^H & = &-\frac{1}{\sqrt{3} F^2} \bigg\{\, \left(M_{a_0}^2-m_{\eta'}^2-m_{\pi^+}^2\right) \bigg[\sqrt{2}\sin\theta(c_{d}^L\sin \varphi_a+c_{d}^H\cos\varphi_a) \nonumber \\
& &  \qquad \qquad \qquad \qquad \qquad \qquad \;  \; +\cos\theta \left((3\alpha_L+2c_{d}^L)\sin\varphi_a
+(3\alpha_H+2c_{d}^H)\cos\varphi_a\right)\, \bigg] \nonumber \\
& & \qquad \qquad + \, 2 \, m_\pi^2(\sqrt{2}\sin\theta+2\cos\theta)(c_{m}^L\sin\varphi_a+c_{m}^H\cos\varphi_a)\,\bigg\} \; ,
\\[3mm]
\mathcal{M}_{a_0^0\to  \eta'\pi^0}^H &=&-\frac{1}{\sqrt{3} F^2} \bigg\{\, \left(M_{a_0}^2-m_{\eta'}^2-m_{\pi^0}^2 \right) \bigg[\sqrt{2}\sin\theta(c_{d}^L\sin \varphi_a+c_{d}^H\cos\varphi_a) \nonumber \\
& & \qquad \qquad \qquad \qquad \qquad \qquad \; \; +\cos\theta \left((3\alpha_L+2c_{d}^L)\sin\varphi_a + (3\alpha_H+2c_{d}^H)\cos\varphi_a\right)\, \bigg] \nonumber \\
& &  \qquad \qquad + \, 2 \, m_\pi^2(\sqrt{2}\sin\theta+2\cos\theta)(c_{m}^L\sin\varphi_a+c_{m}^H\cos\varphi_a)\,\bigg\} \; ,  \nonumber
\\[3mm]
\mathcal{M}_{a_0^+\to  K^+{\overline K}^0}^H & = & -\frac{1}{F^2} \, \bigg( \left(M_{a_0}^2-m_{K^+}^2-m_{K^0}^2\right)(c_{d}^L\sin \varphi_a+c_{d}^H\cos\varphi_a)
+2\,m_K^2(c_{m}^L\sin\varphi_a+c_{m}^H\cos\varphi_a )\bigg)\; , \nonumber
\\[3mm]
\mathcal{M}_{a_0^0\to  K^+K^-}^H & = & -\frac{1}{\sqrt{2} F^2} \bigg( \left(M_{a_0}^2-2m_{K^+}^2 \right)(c_{d}^L\sin \varphi_a+c_{d}^H\cos\varphi_a)
+2\,m_K^2(c_{m}^L\sin\varphi_a+c_{m}^H\cos\varphi_a ) \bigg)\; , \nonumber
\\[3mm]
\mathcal{M}_{a_0^0\to  K^0{\overline K}^0}^H & = & -\frac{1}{\sqrt{2} F^2} \bigg( \left(M_{a_0}^2-2m_{K^0}^2\right)(c_{d}^L\sin \varphi_a+c_{d}^H\cos\varphi_a) +2\,m_K^2(c_{m}^L\sin\varphi_a+c_{m}^H\cos\varphi_a)\bigg) \; , \nonumber
\end{eqnarray}
\begin{eqnarray}
\mathcal{M}_{{K_0^*}^+\to  K^+\pi^0}^H & = & -\frac{1}{ \sqrt{2}F^2} \bigg\{(c_{d}^L\sin \varphi_k+c_{d}^H\cos\varphi_k)\,(M_{{{K}_0^*}^+}^2-m_{K^+}^2-m_{\pi^0}^2)\nonumber\\
&& \qquad \qquad +(c_{m}^L\sin \varphi_k+c_{m}^H\cos\varphi_k)\,(m_K^2+m_\pi^2) \bigg\} \; , \nonumber
\\[3mm]
\mathcal{M}_{{K_0^*}^+\to  K^0\pi^+}^H & = & -\frac{1}{ F^2} \bigg\{(c_{d}^L\sin \varphi_k+c_{d}^H\cos\varphi_k)\,(M_{{{K}_0^*}^+}^2-m_{K^0}^2-m_{\pi^+}^2)\nonumber\\
&& \qquad \qquad +(c_{m}^L\sin \varphi_k+c_{m}^H\cos\varphi_k)\,(m_K^2+m_\pi^2) \bigg\} \; , \nonumber
\\[3mm]
\mathcal{M}_{{K_0^*}^0\to  K^0\pi^0}^H & = & -\frac{1}{ \sqrt{2}F^2} \bigg\{(c_{d}^L\sin \varphi_k+c_{d}^H\cos\varphi_k)\,(M_{{{K}_0^*}^+}^2-m_{K^+}^2-m_{\pi^0}^2)\nonumber\\
&& \qquad \qquad +(c_{m}^L\sin \varphi_k+c_{m}^H\cos\varphi_k)\,(m_K^2+m_\pi^2) \bigg\} \; , \nonumber
\\[3mm]
\mathcal{M}_{{K_0^*}^0\to  K^+\pi^-}^H & = &-\frac{1}{ F^2} \bigg\{(c_{d}^L\sin \varphi_k+c_{d}^H\cos\varphi_k)\,(M_{{{K}_0^*}^+}^2-m_{K^0}^2-m_{\pi^+}^2)\nonumber\\
&& \qquad \qquad +(c_{m}^L\sin \varphi_k+c_{m}^H\cos\varphi_k)\,(m_K^2+m_\pi^2) \bigg\} \; ,
\\[3mm]
\mathcal{M}_{{K_0^*}^+\to  K^+\eta}^H & = &-\frac{1}{ 2\sqrt{3}F^2} \bigg\{\,(M_{{{K}_0^*}^+}^2-m_{K^+}^2-m_{\eta}^2)\bigg[-\sqrt{2}\cos\theta(c_{d}^L\sin \varphi_k+c_{d}^H\cos\varphi_k) \nonumber\\
&&\,\;\;\;\;\;\;\;\;\;\;\;\;\,-2\sin\theta\left(\,(3\alpha_L+2c_{d}^L)\sin\varphi+(3\alpha_H+2c_{d}^H)\cos\varphi\right)\,\bigg]\nonumber\\
&&\,\;\;\;\;\;\;\;\;\;\;\;\;\,+(c_{m}^L\sin\varphi_k+c_{m}^H\cos\varphi_k)\,\bigg[3\sqrt{2}\cos\theta~m_\pi^2-5\sqrt{2}\cos\theta~m_K^2-8\sin\theta~m_K^2\bigg]\,	\bigg\} \; , \nonumber \\[3mm]
\mathcal{M}_{{K_0^*}^0\to  K^0\eta}^H & = &-\frac{1}{ 2\sqrt{3}F^2} \bigg\{\,(M_{{{K}_0^*}^0}^2-m_{K^0}^2-m_{\eta}^2) \bigg[-\sqrt{2}\cos\theta(c_{d}^L\sin \varphi_k+c_{d}^H\cos\varphi_k) \nonumber\\
&&\,\;\;\;\;\;\;\;\;\;\;\;\;\,-2\sin\theta\left(\,(3\alpha_L+2c_{d}^L)\sin\varphi+(3\alpha_H+2c_{d}^H)\cos\varphi\right)\,\bigg]\nonumber\\
&&\,\;\;\;\;\;\;\;\;\;\;\;\;\,+(c_{m}^L\sin\varphi_k+c_{m}^H\cos\varphi_k)\,\bigg[3\sqrt{2}\cos\theta~m_\pi^2-5\sqrt{2}\cos\theta~m_K^2-8\sin\theta~m_K^2\bigg]\,\bigg\} \; . \nonumber
\end{eqnarray}
Those for the decays of the $S_L$ multiplet are:
\begin{eqnarray}
\mathcal{M}_{a_0^+\to  \eta\pi^+}^L & = &-\frac{1}{\sqrt{3} F^2} \bigg\{\, (M_{a_0}^2-m_\eta^2-m_{\pi^+}^2) \bigg[\sqrt{2}\cos\theta(c_{d}^L\cos\varphi_a-c_{d}^H\sin\varphi_a) \nonumber \\
& & \qquad \qquad \qquad \qquad \qquad \qquad \; \;-\sin\theta \left((3\alpha_L+2c_{d}^L)\cos\varphi_a
-(3\alpha_H+2c_{d}^H)\sin\varphi_a\right)\,\bigg] \nonumber \\
& & \qquad \qquad + \, 2 \, m_\pi^2(\sqrt{2}\cos\theta-2\sin\theta)(c_{m}^L\cos\varphi_a-c_{m}^H\sin\varphi_a)\,\bigg\} \,,\nonumber
\\[3mm]
\mathcal{M}_{a_0^0\to  \eta\pi^0}^L & = &-\frac{1}{\sqrt{3} F^2} \bigg\{\, (M_{a_0}^2-m_\eta^2-m_{\pi^0}^2) \bigg[\sqrt{2}\cos\theta(c_{d}^L\cos\varphi_a-c_{d}^H\sin\varphi_a) \nonumber \\
& & \qquad \qquad \qquad \qquad \qquad \qquad \; \; -\sin\theta \left((3\alpha_L+2c_{d}^L)\cos\varphi_a
-(3\alpha_H+2c_{d}^H)\sin\varphi_a\right)\,\bigg] \nonumber \\
& & \qquad \qquad + \, 2 \, m_\pi^2(\sqrt{2}\cos\theta-2\sin\theta)(c_{m}^L\cos\varphi_a-c_{m}^H\sin\varphi_a)\,\bigg\} \,, \nonumber
\\[3mm]
\mathcal{M}_{a_0^+\to  \eta'\pi^+}^L & = &-\frac{1}{\sqrt{3} F^2} \bigg\{\, (M_{a_0}^2-m_{\eta'}^2-m_{\pi^+}^2) \bigg[\sqrt{2}\sin\theta(c_{d}^L\cos \varphi_a-c_{d}^H\sin\varphi_a) \nonumber \\
& & \qquad \qquad \qquad \qquad \qquad \qquad \; \; +\cos\theta \left((3\alpha_L+2c_{d}^L)\cos\varphi_a
 -(3\alpha_H+2c_{d}^H)\sin\varphi_a\right)\,\bigg] \nonumber \\
 & & \qquad \qquad + \, 2 \, m_\pi^2(\sqrt{2}\sin\theta+2\cos\theta)(c_{m}^L\cos\varphi_a-c_{m}^H\sin\varphi_a)\,\bigg\} \,, \nonumber
\\[3mm]
\mathcal{M}_{a_0^0\to  \eta'\pi^0}^L & = &-\frac{1}{\sqrt{3} F^2} \bigg\{\, (M_{a_0}^2-m_{\eta'}^2-m_{\pi^0}^2)\bigg[\sqrt{2}\sin\theta(c_{d}^L\cos \varphi_a-c_{d}^H\sin\varphi_a) \nonumber \\
& & \qquad \qquad \qquad \qquad \qquad \qquad \; \; +\cos\theta \left((3\alpha_L+2c_{d}^L)\cos\varphi_a
-(3\alpha_H+2c_{d}^H)\sin\varphi_a\right)\,\bigg] \nonumber \\
& & \qquad \qquad + \, 2 \, m_\pi^2(\sqrt{2}\sin\theta+2\cos\theta)(c_{m}^L\cos\varphi_a-c_{m}^H\sin\varphi_a)\,\bigg\} \; , \nonumber
\\[3mm]
\mathcal{M}_{a_0^+\to  K^+{\overline K}^0}^L & = & -\frac{1}{F^2} \, \bigg( (M_{a_0}^2-m_{K^+}^2-m_{K^0}^2)(c_{d}^L\cos\varphi_a-c_{d}^H\sin\varphi_a) \nonumber \\
& & \qquad \; \; +2\,m_K^2(c_{m}^L\cos\varphi_a-c_{m}^H\sin\varphi_a) \bigg)\; , \nonumber
\\[3mm]
\mathcal{M}_{a_0^0\to  K^+K^-}^L & = & -\frac{1}{\sqrt{2} F^2} \bigg( (M_{a_0}^2-m_{K^+}^2-m_{K^0}^2)(c_{d}^L\cos\varphi_a-c_{d}^H\sin\varphi_a)
\nonumber \\
& & \qquad \qquad \; \;  +2\,m_K^2(c_{m}^L\cos\varphi_a-c_{m}^H\sin\varphi_a) \bigg)\; , \nonumber
\\[3mm]
\mathcal{M}_{a_0^0\to  K^0{\overline K}^0}^L & = & -\frac{1}{\sqrt{2} F^2} \bigg( (M_{a_0}^2-2 m_{K^0}^2)(c_{d}^L\cos\varphi_a-c_{d}^H\sin\varphi_a) \nonumber \\
& & \qquad \qquad \; \; +2\,m_K^2(c_{m}^L\cos\varphi_a-c_{m}^H\sin\varphi_a) \bigg) \;  ,
\end{eqnarray}
\begin{eqnarray}
\mathcal{M}_{{K_0^*}^+\to  K^+\pi^0}^L & = & -\frac{1}{ \sqrt{2}F^2} \bigg\{(c_{d}^L\cos \varphi_k-c_{d}^H\sin\varphi_k)\,(M_{{K_0^*}^+}^2-m_{K^+}^2-m_{\pi^0}^2)\nonumber\\
&& \qquad \qquad  +(c_{m}^L\cos \varphi_k-c_{m}^H\sin\varphi_k)\,(m_K^2+m_\pi^2) \bigg\} \; , \nonumber
\\[3mm]
\mathcal{M}_{{K_0^*}^+\to  K^0\pi^+}^L & = & -\frac{1}{ F^2} \bigg\{(c_{d}^L\cos \varphi_k-c_{d}^H\sin\varphi_k)\,(M_{{K_0^*}^+}^2-m_{K^0}^2-m_{\pi^+}^2)\nonumber\\
&& \qquad  +(c_{m}^L\cos \varphi_k-c_{m}^H\sin\varphi_k)\,(m_K^2+m_\pi^2)\bigg\} \; , \nonumber
\\[3mm]
\mathcal{M}_{{K_0^*}^0\to  K^0\pi^0}^L & = & -\frac{1}{ \sqrt{2}F^2} \bigg\{(c_{d}^L\cos \varphi_k-c_{d}^H\sin\varphi_k)\,(M_{{K_0^*}^+}^2-m_{K^+}^2-m_{\pi^0}^2)\nonumber\\
&& \qquad \qquad +(c_{m}^L\cos \varphi_k-c_{m}^H\sin\varphi_k)\,(m_K^2+m_\pi^2) \big\} \; , \nonumber
\\[3mm]
\mathcal{M}_{{K_0^*}^0\to  K^+\pi^-}^L & = &-\frac{1}{ F^2} \bigg\{(c_{d}^L\cos \varphi_k-c_{d}^H\sin\varphi_k)\,(M_{{K_0^*}^+}^2-m_{K^0}^2-m_{\pi^+}^2)\nonumber\\
&& \qquad +(c_{m}^L\cos \varphi_k-c_{m}^H\sin\varphi_k)\,(m_K^2+m_\pi^2) \bigg\} \; ,
\\[3mm]
\mathcal{M}_{{K_0^*}^+\to  K^+\eta}^L & = &-\frac{1}{ 2\sqrt{3}F^2} \big\{\,(M_{{K_0^*}^+}^2-m_{K^+}^2-m_{\eta}^2)\bigg[-\sqrt{2}\cos\theta(c_{d}^L\cos \varphi_k-c_{d}^H\sin\varphi_k) \nonumber\\
&&\,\;\;\;\;\;\;\;\;\;\;\;\;\,+2\sin\theta\left(\,(3\alpha_H+2c_{d}^H)\sin\varphi-(3\alpha_L+2c_{d}^L)\cos\varphi\right)\,\bigg]\nonumber\\
&&\,\;\;\;\;\;\;\;\;\;\;\;\;\,+(c_{m}^L\cos\varphi_k-c_{m}^H\sin\varphi_k)\,\bigg[3\sqrt{2}\cos\theta~m_\pi^2-5\sqrt{2}\cos\theta~m_K^2-8\sin\theta~m_K^2\bigg]\,\bigg\} \; , \nonumber \\[3mm]
\mathcal{M}_{{K_0^*}^0\to  K^0\eta}^L & = &-\frac{1}{ 2\sqrt{3}F^2} \bigg\{\,(M_{{K_0^*}^0}^2-m_{K^0}^2-m_{\eta}^2)\bigg[-\sqrt{2}\cos\theta(c_{d}^L\cos \varphi_k-c_{d}^H\sin\varphi_k) \nonumber\\
&&\,\;\;\;\;\;\;\;\;\;\;\;\;\,+2\sin\theta\left(\,(3\alpha_H+2c_{d}^H)\sin\varphi-(3\alpha_L+2c_{d}^L)\cos\varphi\right)\,\bigg]\nonumber\\
&&\,\;\;\;\;\;\;\;\;\;\;\;\;\,+(c_{m}^L\cos\varphi_k-c_{m}^H\sin\varphi_k)\,\bigg[3\sqrt{2}\cos\theta~m_\pi^2-5\sqrt{2}\cos\theta~m_K^2-8\sin\theta~m_K^2 \bigg]\,\bigg\} \; . \nonumber
\end{eqnarray}

\subsection{\B{I = O} decays} \label{ssap:singlet}
In the following amplitudes, $i=1,2,3$ and $f_1 \equiv f_0(1370)$, $f_2 \equiv f_0(1500)$ and $f_3 \equiv f_0(1710)$:
\begin{eqnarray}
\label{eq:fii9}
\mathcal{M}_{f_i\to \pi^+\pi^-, \pi^0 \pi^0} & = & \frac{M_{f_i}^2-2 m_{\pi}^2}{3 \, F^2} \, \left[\sqrt{6} \,  c_{d}^H \, \left( a_{i1} + \sqrt{2} \,  a_{i2} \right) \, + \, 6 \,  \sqrt{3} \, \beta_H \, a_{i2} \, + \,  6 \, c_d' \,  a_{i3} \right]  \nonumber \\[3mm]
&& + \, 4 \, \frac{m_{\pi}^2}{F^2} \left[ \frac{c_{m}^H}{\sqrt{6}} \, \left( a_{i1}\,  + \, \sqrt{2} \, a_{i2} \right) \, + \, c_{m}' \, a_{i3} \right] \nonumber \\[3mm]
\nonumber \\[3mm]
\mathcal{M}_{f_i\to  K^+K^-, K^0 \overline{K^0}} & = & \frac{M_{f_i}^2-2 m_{K}^2}{6 \, F^2} \, \left[\sqrt{6} \,  c_{d}^H \, \left( - \,  a_{i1} \, + \, 2 \, \sqrt{2} \,  a_{i2} \right) \, + \, 12 \,  \sqrt{3} \, \beta_H \, a_{i2} \, + \,  12 \, c_d' \,  a_{i3} \right]  \nonumber \\[3mm]
&& + \, \frac{m_{K}^2}{3 \, F^2} \left[ \sqrt{6} \,  c_{m}^H \left( - \, a_{i1}\,  + \, 2 \, \sqrt{2} \, a_{i2} \right) \, + \, 12 \,  c_{m}' \, a_{i3} \right] , \nonumber \\
\end{eqnarray}
\begin{eqnarray}
\label{eq:fii19}
\mathcal{M}_{f_i\to  \eta\eta} & = & \frac{M_{f_i}^2-2 m_{\eta}^2}{6 \, F^2} \, \left[ -\,\sqrt{6} \, c_{d}^H \, a_{i1} \, + \, 12 \, c_d' \, a_{i3} \, + \,
2 \, \sqrt{3} \, a_{i2} \left( 3 \alpha_H + 6 \beta_H + 2 c_{d}^H+ 9 \gamma_H \right) \, + \, 18 \, \gamma_H \, a_{i3} \right. \nonumber \\[3mm]
& & \left. \qquad \qquad \qquad - \, \cos 2 \theta \, \left[ \sqrt{6} \, c_{d}^H \, a_{i1} \, + \, 6 \, \sqrt{3} \, a_{i2} \, ( \,\alpha_H \, + \,  3 \, \gamma_H) \, + \, 18 \,  a_{i3} \, \gamma' \right] \right. \nonumber \\[3mm]
& & \left. \qquad \qquad \qquad - \, 2 \, \sqrt{3} \, \sin 2 \theta \, \left( 3 \, \alpha_H \, + 2 \, c_{d}^H \right) \, a_{i1} \right] \nonumber \\[3mm]
& & + \, \frac{1}{9 \, F^2} \left[ \sqrt{6} \, c_{m}^H \, a_{i1} \, \left[ 3 \, ( 3 m_{\pi}^2 - 4 m_K^2 ) \, + \, (m_{\pi}^2-4 m_K^2 ) ( \cos 2 \theta \, + \, 2 \, \sqrt{2} \,
\sin 2 \theta ) \right] \right. \nonumber \\[3mm]
& & \qquad \qquad \left. + \, 4 \, \left( \sqrt{3}  \, c_{m}^H \, a_{i2} + 3 c_m' \, a_{i3} \right) \left( 3 \, m_K^2 \, + \, (m_K^2 - m_{\pi}^2) ( \cos 2 \theta \, + \, 2 \, \sqrt{2} \, \sin 2 \theta ) \right) \right] \nonumber \\[3mm]
\mathcal{M}_{f_i\to  \eta\eta'} & = & \frac{M_{f_i}^2 - m_{\eta}^2 - m_{\eta'}^2}{6 \, F^2} \, \left[ 2 \, \sqrt{3} \, (3 \, \alpha_H \, + \, 2 \, c_{d}^H ) \, a_{i1} \, \cos 2 \theta \, -  \right. \nonumber \\[3mm]
& & \qquad \qquad \qquad \qquad \left. - \, \left( \sqrt{6} \, c_{d}^H \, a_{i1} \, + \, 6 \, \sqrt{3} \, (\alpha_H + 3 \gamma_H) \, a_{i2} \, + \, 18 \, a_{i3} \, \gamma' \right) \sin 2 \theta \, \right] \nonumber \\[3mm]
&& + \, \frac{1}{9 \, F^2} \, \left( 2 \, \sqrt{2} \cos 2 \theta \, - \, \sin 2 \theta \right) \left[ \sqrt{6} \, c_{m}^H \, a_{i1} \, ( 4 m_K^2 - m_{\pi}^2)
\right. \nonumber \\[3mm]
&& \qquad \qquad \qquad \qquad \qquad \qquad \qquad\left.  - \, 4 (m_K^2 - m_{\pi}^2) \left( \sqrt{3} \, c_{m}^H \, a_{i2} \, + \, 3 \, c_m' \, a_{i3} \right) \right] .
\end{eqnarray}
Here $a_{ij}$ are the matrix elements of the $A$ matrix in Eq.~(\ref{eq:matrixa}).
In Eq.~(\ref{eq:fii19}), $\theta$ is the $\eta-\eta'$ mixing angle defined by:
\begin{equation}
\label{eq:etamix}
\left( \begin{array}{c} \eta \\ \eta' \end{array} \right) =  \left( \begin{array}{cc} \cos \theta & - \sin \theta \\ \sin \theta & \cos \theta \end{array} \right) \, \left( \begin{array}{c} \eta_8 \\ \eta_0 \end{array} \right) .
\end{equation}

\setcounter{equation}{0}
\setcounter{table}{0}
\section{Final State Interactions in $\mathbold{f_i \rightarrow PP}$ decays} \label{app:fsi}
We know that $I=0, S=0$ amplitudes have large FSI effects. Unfortunately we only have reliable information on the $\pi \pi$ and $K \overline{K}$ phase-shifts. Hence we can only consider the FSI effects in the decays with those final states.
We would expect that $I=1$ or $I=1/2$ final states should be less affected and, therefore, we will consider only $f_i \rightarrow PP$ decays ($i=1,2,3$ as in Appendix~\ref{app:damp}), with $P=\pi,K$.
\par
Following Refs.~\cite{Smith:1998nu,Suzuki:2007je},  we can parameterize:
\begin{equation} \label{eq:fs1}
\left( \begin{array}{c}
        \mathcal{M}_{f_i \rightarrow \pi \pi} \\
        \mathcal{M}_{f_i \rightarrow K \overline{K}}
        \end{array} \right)^{\mbox{\tiny{FSI}}}
        \; = \;
        \sqrt{S}
\left( \begin{array}{c}
        \mathcal{M}_{f_i \rightarrow \pi \pi} \\
        \mathcal{M}_{f_i \rightarrow K \overline{K}}
        \end{array} \right)^{\mbox{\tiny{bare}}}     \, ,
\end{equation}
where
\begin{equation} \label{eq:fs2}
\sqrt{S} \; = \; {\cal O}^{T} \, \sqrt{S_{\mbox{\tiny{diag}}}} \, {\cal O} \, ,
\end{equation}
with
\begin{eqnarray} \label{eq:fs3}
S_{\mbox{\tiny{diag}}} \, = \, \left( \begin{array}{cc}
                                       e^{2 i \delta_{\pi \pi}^{I=0}} & 0 \\
                                       0 & e^{2 i \delta_{K \overline{K}}^{I=0}}
                                       \end{array} \right) \; \; \; \; \; \;  &,& \; \; \;  \; \; \;
{\cal O} \, = \, \left( \begin{array}{cc}
                        \cos \omega & \sin \omega \\
                        - \sin \omega & \cos \omega
                        \end{array} \right) .
\end{eqnarray}
Here $\omega$ should be a new parameter to fit. For the phase-shifts we will only need $\delta_{\pi \pi}^{I=0}(M_{f_i}^2)$ and
$\delta_{K \overline{K}}^{I=0}(M_{f_i}^2)$, because in two-body decays always $s=M^2$ being $M$ the mass of the decaying particle. The phase shifts are given by the extended K-matrix fit following \cite{Dai:2014lza,Dai:2014zta}, up to $1.8 \, \mbox{GeV}$. We will consider the results in Table~\ref{tab:phasesh}.
\begin{table}
\begin{center}
\renewcommand{\arraystretch}{1.5}
\begin{tabular}{|c|c|c|}
\hline
\multicolumn{1}{|c|}{Energy (GeV)} &
\multicolumn{1}{|c|}{$\delta_{\pi \pi}^{I=0}$ (Deg)} &
\multicolumn{1}{|c|}{$\delta_{\overline{K}K}^{I=0}$ (Deg)} \\
\hline
\hline
$ 1.395 $ &   $308.05$ &  $-71.46$ \\
$ 1.504 $ &   $340.18$ &  $-78.92$ \\
$ 1.720 $ &   $373.59$ &  $-107.20$ \\
\hline
\multicolumn{3}{c}{}
\end{tabular}
\end{center}
\vspace*{-1cm}
\caption{\label{tab:phasesh}
Phase shifts for the FSI interactions in $f_i \rightarrow \pi \pi, \, \overline{K}K$ decays. Data from
\cite{Dai:2014lza,Dai:2014zta}.}
\end{table}

\providecommand{\href}[2]{#2}\begingroup\raggedright\endgroup


\end{document}